\documentclass[11pt]{article}   	
\usepackage{graphics, subfigure, theorem, times, amsfonts, amsmath, amssymb, cite}
\usepackage{pgfplots}
\usepackage{color}
\usepackage{hyperref,url}
\definecolor{darkblue}{rgb}{0,0.08,0.4} 
\definecolor{darkred}{rgb}{0.4,0.08,0} 
\input{mysymbol.sty}
\usepackage[margin=0.75in]{geometry}

\newtheorem{theorem}{Theorem}
\newtheorem{proposition}{Proposition}
\newtheorem{corollary}{Corollary}
\newtheorem{lemma}{Lemma}
\newtheorem{definition}{Definition}
\newtheorem{example}{Example}

\newenvironment{proof}{\par
   \noindent \textit{Proof:} \itshape}{}

\begin{document}
\title{Disease dynamics on a network game: a little empathy goes a long way} 
\author{\hspace{0in}\fontsize{13}{13}\selectfont {Ceyhun Eksin}\thanks{School of Biology, Georgia Institute of Technology, Atlanta, GA }\ \ \thanks{\fontsize{9}{9}\selectfont School of Electrical and Computer Engineering, Georgia Institute of Technology, Atlanta, GA}\ \ \and
Jeff S. Shamma\thanks{Computer, Electrical and Mathematical Sciences and Engineering, King Abdullah University of Science and Technology (KAUST) Thuwal, Saudi Arabia} \and
{Joshua S. Weitz}\thanks{\fontsize{9}{9}\selectfont  School of Physics, Georgia Institute of Technology, Atlanta, GA}$\ \ ^{*}$}

\date{\let\thefootnote\relax\footnote{This work is supported by Army Research Office  grant \#W911NF-14-1-0402, and supported in part by KAUST. The authors thank K. Paarporn (Georgia Inst. Tech.) and J. W. Glasser (Center for Disease Control (CDC)) for their comments. }
\let\thefootnote\relax\footnote{To whom correspondence should be addressed. E-mail: ceyhuneksin@gatech.edu, jsweitz@gatech.edu}}


\maketitle 

\begin{abstract}
Individuals change their behavior during an epidemic in response to whether they and/or those they interact with are healthy or sick. Healthy individuals are concerned about contracting a disease from their sick contacts and may utilize protective measures. Sick individuals may be concerned with spreading the disease to their healthy contacts and adopt preemptive measures. Yet, in practice both protective and preemptive changes in behavior come with costs. This paper proposes a stochastic network disease game model that captures the self-interests of individuals during the spread of a susceptible-infected-susceptible (SIS) disease where individuals react to current risk of disease spread, and their reactions together with the current state of the disease stochastically determine the next stage of the disease. We show that there is a critical level of concern, i.e., empathy, by the sick individuals above which disease is  eradicated fast. Furthermore, we find that if the network and disease parameters are above the epidemic threshold, the risk averse behavior by the healthy individuals cannot eradicate the disease without the preemptive measures of the sick individuals. This imbalance in the role played by the response of the infected versus the susceptible individuals in disease eradication affords critical policy insights.
\end{abstract}



Infectious diseases change the social interaction patterns in the society they impact. During the Ebola outbreak, many studies pointed to changes in social customs, e.g., switch to safe burial methods from traditional ceremonial burials, playing a critical role in impeding disease spread \cite{Pandey_et_al_2014}. Similar behavioral responses played important roles in modifying disease spread in other pandemics, e.g., wearing protective masks during the SARS pandemic \cite{Chowell_et_al_2003,Lau_et_al_2004,Pang_et_al_2003}, decrease in unprotected sex when STD is at high levels \cite{Hethcote_Yorke_1984,Hyman_Li_1997} or covering one's own cough and staying at home if sick during a flu pandemic \cite{Nelson_2004,Jones_Salathe_2009,Steelfisher_et_al_2010}. These responses to prevalence of the disease are efforts to preempt disease spread by the infected and the susceptible individuals in the population.

In many infectious diseases infected and susceptible individuals have to be in contact for disease transmission. Accordingly, there has been a surge of interest on disease spread models in which a contact network determines the subset of individuals that can be infected by an infected individual \cite{Volz_Meyers_2007,Meyers_et_al_2005,Bansal_et_al_2007,Volz_Meyers_2009,Mieghem_et_al_2009,Volz_et_al_2011}. These studies were influential in relating network structural properties to epidemic thresholds and in revealing the limits to inferences made by models that assume homogeneous mixing.

The rate at which individuals meet with their contacts changes depending on the individual preemptive measures during the course of a disease \cite{Bauch_Galvani_2013}. Consequently, a number of dynamic models have been developed to assess the effects individual preemptive measures have on infectious disease spread over networks \cite{Perra_et_al_2011,Mbah_et_al_2012,Paarporn_et_al_15,Funk_et_al_2009}. These models couple behavior and disease dynamics. That is, the state of the disease and the contact network determine the preemptive measures of the individuals which then affect the disease spread. Preemptive measures in these models, which are in the form of social distancing or rewiring of transmissive links, are results of simple heuristics that approximate the decision-making of healthy individuals. These heuristic based decision-making algorithms are intended to be approximations of decisions made by self-interested individuals. 

When individuals act according to their selfish interests, they would compare the inherent costs of these preemptive measures with the risks of disease contraction. However, the actions of other individuals also affect the risk of infection. Game theory provides a means to consider how individuals make rational decisions by reasoning strategically about the decisions of others. Recent epidemiological models with game theoretic individual decision-making either consider one-shot rational decisions of all susceptible individuals at the beginning, e.g., vaccinate or not, social distance or not, that determines the course of the disease \cite{Bauch_et_al_03,Bauch_Earn_2004,Molina_Earn_2015,Perisic_Bauch_2009,Omic_et_al_2009,Shim_et_al_2012}, or use bounded rational heuristics for repeated decision-making \cite{Enright_Kao_2015,Zhang_et_al_13,Cornforth_et_al_2011,Reluga_2010} (see \cite{Wang_et_al_2015} for a recent extensive review).

Here, we consider individuals---susceptible and infected--- making daily rational decisions on preemptive measures, e.g., social distancing, staying home from school/work, wearing protective masks, based on the current risks of disease spread for a susceptible-infected-susceptible (SIS) infection. In particular, a healthy individual compares the cost of protection measures with its current risk of infection. This means a healthy individual can forgo any protective measure (free-ride) if it perceives its sick contacts are taking the utmost preemptive measures. However, at the same time, a sick individual compares the cost of preemptive measures with the current risk of spreading the disease to its healthy neighbors. This means sick individuals have to reason strategically about the decisions of their healthy neighbors who reason about the decisions of their sick neighbors. This sets up a daily game among healthy and sick individuals. The daily rational measures taken by both the healthy and the sick as a result of the \emph{network disease game} set the probabilities of disease contractions which in turn stochastically determine the status of the disease in the following day. 

Using this model, we explore the interrelationship among contact network structure, rational daily decisions and SIS disease dynamics. Specifically, we provide analytical bounds for the initial spread of the disease from a single infected individual based on selection of the initial infected individual, network degree distribution, disease infection and healing rates, and the relative weight (\emph{empathy}) that sick individuals have on disease spread versus cost of preemptive measures. These bounds show that increasing \emph{empathy} of the sick individuals stops the initial disease spread. We also show that these bounds are good indicators of disease eradication starting from any initial level of infection, not just a single infected individual. Moreover, we confirm past results in \cite{Paarporn_et_al_15,Funk_et_al_2009} by showing that susceptible individuals cannot by themselves quickly eradicate the disease no matter how risk averse they are if the empathy factor is zero. Yet, for any positive empathy by the sick individuals, we can find a critical risk averseness factor value that can help eradicate the disease. These results imply that a little preemptive action by the sick individuals is crucial in fast eradication of an epidemic.

\section{Model}

We consider stochastic SIS disease dynamics where an individual $i$ in the population $\ccalN:=\{1,\dots,n\}$ is either susceptible ($s_i(t) =0$) or infected ($s_i(t) =1$) at any given time $t=1, 2,\dots$. If the individual is susceptible at time $t$, it gets infected at time $t+1$ with probability $p^i_{01}(t) := \mathbb{P}[s_i(t+1) = 1 | s_i(t) = 0]$. If the individual is sick at time $t$, it becomes susceptible at time $t+1$ with probability $p^i_{10}(t) := \mathbb{P}[s_i(t+1) = 0 | s_i(t) = 1]$. These transition probabilities define a Markov chain for the  disease dynamics of individual $i \in \ccalN$ as follows
\begin{align} \label{eq_markov_chain}
s_i(t+1) = \begin{cases}
1 \ \mbox{ with prob. } p^i_{01}(t) &   \mbox{ if } s_i(t) = 0\\
0 \ \mbox{ with prob. } 1-p^i_{01}(t)& \mbox{ if } s_i(t) = 0 \\
1 \ \mbox{ with prob. } 1-p^i_{10}(t)&  \mbox{ if } s_i(t) = 1\\
0 \ \mbox{ with prob. } p^i_{10}(t) &   \mbox{ if } s_i(t) = 1.
\end{cases}
\end{align}

\begin{figure}[t]
\centering
\includegraphics{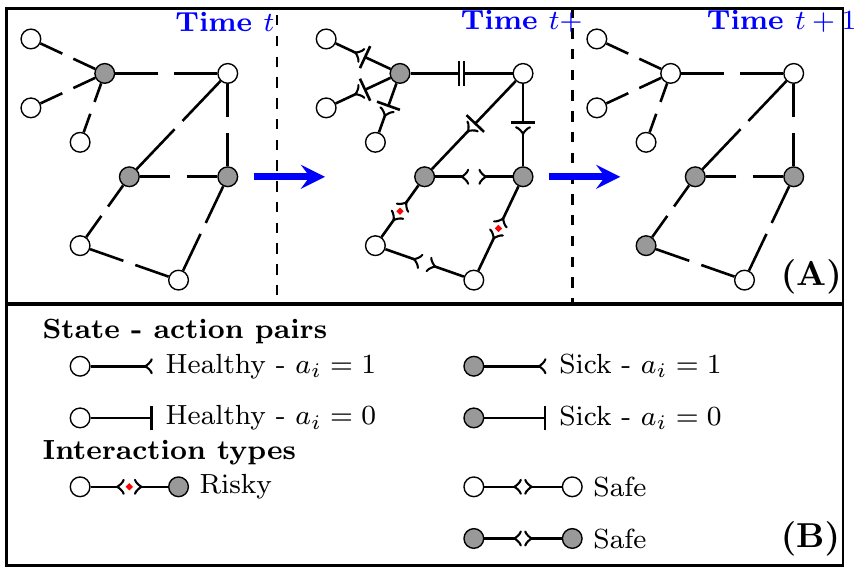} 
\caption{SIS Markov chain dynamics. We show one step of the Markov chain dynamics in {\bf (A)} where circles denote individuals who are healthy (open) or sick (shaded). {The edges denote contacts, where the actions are indicated by the edge-ends types, social distancing $|$ and interaction $\succ$}. Given the state of the disease and contact network at time t, individuals decide to take preemptive measures or not at time $t+$ which determines the state of each individual at time $t+1$ according to the Markov chain dynamics in \eqref{eq_markov_chain}. If $a_i =1$, individual $i$ does not take any preemptive measures. If  $a_i=0$, $i$ self-quarantines itself reducing any risk of disease contraction or spread to zero. A healthy individual can only contract the disease from an interaction with a neighbor if and only if its neighbor is infected and neither of the two self-quarantines. These interactions are marked by a red dot in the middle figure. We enumerate each pair of state and action in {\bf (B)}. 
\label{fig:example_sis}}
\end{figure}

A susceptible individual can only contract the disease in the next time step if in contact with an infected individual. We define the set of contacts of each individual by a contact network $\ccalG$ with node set $\ccalN$ and edge set $\ccalE$ -- see Fig.\ \ref{fig:example_sis}(A) for an example. The contact neighborhood of individual $i$ is $\ccalN_i := \{j: (i,j)\in \ccalE\}$. The chance of a susceptible individual ($s_i(t) = 0$) contracting the disease from a neighboring infected contact ($s_j(t) = 1$) depends on the infection probability of the disease $\beta \in (0,1)$, action of $i$ $a_i(t)$ belonging to the unit interval $[0,1]$, and the contact's action $a_j(t) \in [0,1]$. In particular, the probability of $i\in\ccalN$ getting infected from individual $j \in \ccalN_i$ is equal to $\beta a_i(t) a_j(t) s_j(t)$. Assuming each local interaction is independent, we have 
\begin{align}
p_{01}^i(t) = 1-\prod_{j\in \ccalN_i} \left(1 - \beta a_i(t) a_j(t) s_j(t)\right).\label{eq:Markov}
\end{align}
Each term inside the product is the probability that the individual $i$ is not infected by its neighbor $j$. The product of all the terms is the probability that the individual is not infected by any one of its interactions. Finally, subtracting the product from one gives the probability that individual $i$ contracts the disease.

The other important event in the SIS disease dynamics in \eqref{eq_markov_chain} is the transition of an infected individual to a susceptible state which is equal to the inherent healing rate of the disease $\delta \in (0,1)$, i.e., $p^i_{10}(t) =\delta$ for $t=1,2,\dots$.

The individual $i$'s action ($a_i(t)$) represents preemptive measures, e.g., wearing a protective mask, or reducing social interaction, that the individual can take at time $t$, where $a_i(t)=0$ means \emph{self-quarantine} and $a_i(t)=1$ means resuming \emph{normal} social interaction with no protective measures. If the actions of all individuals are equal to one, $a_i(t) =1$ for all times, then the model recovers the disease spread models over contact networks that do not include individual behavior response \cite{Mieghem_et_al_2009}. The Markov chain disease dynamics in \eqref{eq_markov_chain} generalizes these models to include preemptive actions ($a_i(t)$ and $a_j(t)$) as variables that affect infection probability $p^i_{01}(t)$ in \eqref{eq:Markov}.

According to the infection probability in \eqref{eq:Markov}, a healthy individual $i$ is able to protect itself from infection by {self-quarantining} ($a_i(t) = 0$). Individual $i$ is also safe from the disease regardless of its action if its infected neighbors  self-quarantine ($a_j(t) = 0$ for $j \in \ccalN_i$) -- see Fig.\ \ref{fig:example_sis}(A) for an example and see Fig.\ \ref{fig:example_sis}(B) for state and action pairs and interaction types. That is, individual $i$ can avoid the disutility of protective measures, i.e., take action $a_i(t) = 1$, if it knows its infected neighbors are taking  preemptive measures to avoid disease spread. However, its neighbors also would like to avoid the disutility of these preemptive measures. Hence individual $i$ has to reason about their preferences to know its neighbors' actions. In the following, we specify the tradeoffs that individuals face between risk of disease spread and cost of preemptive measures, and then present a game theoretic framework that describes how individuals reason about each others' actions and determine their actions.

\subsection{Individual preferences: a bilinear game}
Individuals determine their actions based on their risk of getting infected or their potential to infect others. If individual $i$ is susceptible ($s_i(t) = 0$), individual $i$ has a concern for contracting the disease proportional to its probability of infection $p^i_{01}(t)$ in the next time step. If individual $i$ is infected ($s_i(t) = 1$), individual $i$ is concerned about infecting others in its neighborhood. This concern is proportional to the probability that $i$ infects $j$ ($\beta a_i a_j $) for $j\in\ccalN_i$. Finally, individual $i$ has an incentive to avoid the cost of preemptive measures, i.e., continue its normal levels of social interaction. A weighted linear combination of these preferences have the following form,
\begin{align} \label{utility_raw}
{c_0 a_i} - {c_1 (1-s_i(t)) p_{01}^i(t)} - {\beta c_2  a_i s_i(t) \sum_{j\in\ccalN_i}  a_j (1-s_j(t))} 
 \end{align}
where $c_0, c_1, c_2$ are positive weights. The first term is the payoff of $i$ from resuming normal activity, the second term is the concern for getting infected which is nonzero only when individual $i$ is susceptible ($s_i(t) = 0$), and the third term is the concern for infecting others which is nonzero only when individual $i$ is infected ($s_i(t) = 1$). We will refer to $c_0$, $c_1$, and $c_2$ as \emph{socialization}, \emph{risk averseness} and \emph{empathy} constants, respectively.  

We use the following lower bound of the preference function defined above in \eqref{utility_raw} as the utility of the individuals (see Appendix A for the derivation),
\begin{align} \label{utility}
 u_i(a_i, \{a_{j}\}_{j\in \ccalN_i}, s(t)) = &
  a_i \beta \bigg({c_0}  -  c_1 (1-s_i(t)) \sum_{j\in \ccalN_i} a_j s_j(t) - { c_2 s_i(t) \sum_{j\in\ccalN_i} a_j (1-s_j(t))}\bigg),
\end{align}
where $s(t) = \{s_1(t),\dots, s_n(t)\}$ represents the state of the disease at time $t$. The terms inside the parentheses have intuitive explanations identical to the three terms in the preference function above in \eqref{utility_raw}. The first term is the socialization payoff, the second term is the cost of risk aversion that increases with the number of infected neighbors that do not take any preemptive measures, and the third term is the cost from the risk of infecting others that increases with the number of healthy neighbors that do not take any protective measures. 

Note that the payoff above is a bilinear function of $a_i$ and $a_j$ for $j \in \ccalN_i$. Maximizing the above utility function given $\{a_{j}\}_{j\in \ccalN_i}$ and $s(t)$ with respect to individual $i$'s action $a_i$, we obtain whether an individual resumes normal activity ($a_i = 1$) or self-quarantines ($a_i = 0$) depending on the sign of expression inside the parantheses. In particular, if this expression is positive, $i$ takes action $a_i = 1$. Otherwise, the action that maximizes the utility is to self-quarantine. 

In reality, the actions of neighboring individuals $\{a_{j}\}_{j\in \ccalN_i}$ are not available to the individual. The payoffs of the neighbors of individual $i$ depend on the actions of their own neighbors, i.e., the actions of other individuals as well as $i$. This means in a connected contact network $G$, payoffs couple the actions of all the individuals. Hence, individuals need to reason about the interaction levels of their neighbors in their decision-making. We model individuals' reasoning using game theory.

%
\section{Results}


\subsection{Stochastic network game: strategic behavior and its existence}

Stochastic games denotes a class of games where the current state determines the payoffs of players, and the actions taken in the current time step determine the state in the following time \cite{Fudenberg_Tirole_1991}. Because the payoffs of individuals in \eqref{utility} depend on their neighbors only, the game we consider is a stochastic network game. A rational model of decision-making in a stochastic game is the Markov perfect equilibrium solution concept \cite[Ch.5.5]{Mailath_Samuelson_2006}. A Markov strategy is where individuals' actions depend on the payoff relevant state $s(t)$. In addition, we assume individuals repeatedly take actions considering their current payoffs only. This equilibrium concept, we term the myopic Markov perfect equilibrium, is formally defined as follows.  
\begin{definition} \label{def:myopic_MPE}
The strategy of individual $i$ at time $t$, $\sigma_{i}$ is a mapping from the state $s(t)$ to its action space $[0,1]$, i.e., $a_i^{*}(t) = \sigma_{i}(s(t)) $. A {myopic Markov perfect equilibrium (MMPE)} strategy profile $\sigma:=\{\sigma_{i}\}_{i\in \ccalN}$ is such that for all $t =1, 2,\dots$, the state $s(t)$ and $i \in \ccalN$ it holds that
\begin{equation} \label{eq:myopic_MPE}
u_i(a_i^{*}(t), \{a_{j}^{*}(t)\}_{j\in \ccalN_i}, s(t)) \geq u_i(a_i, \{a_{j}^{*}(t)\}_{j\in \ccalN_i}, s(t)), 
\end{equation}
for any $a_i \in [0,1]$ and the state evolves according to the Markov chain in \eqref{eq_markov_chain}.
\end{definition}

An MMPE strategy profile at time $t$ is a Nash equilibrium of the stage game at time $t$ where the stage game is defined by the payoffs given the state $s(t)$. That is, no individual has a preferable unilateral deviation to another action that strictly increases its stage payoff at any stage $t$. When the Markov strategy profile is independent of time, i.e., it only depends on the state, the strategy is stationary. The equilibrium definition given above implies that an MMPE strategy profile is stationary. The assumption of myopic strategies implies that individuals do not weigh their future risks of infection or infecting other individuals in their current decision-making. This is a reasonable initial assumption considering individuals have local information about the state of the disease and the computational complexity of considering future states of the disease during an epidemic. 

We defined the MMPE strategy profile such that each individual's action at each stage is determined by the state of the disease. That is, individuals' strategies are degenerate distributions on the action space, known as pure strategy equilibria \cite{Fudenberg_Tirole_1991}. 
In Appendix B, we show that there exists at least one such strategy profile for the bilinear game in \eqref{utility}. Our proof of existence is constructive and it yields an algorithm that computes an MMPE strategy profile in finite time. This  demonstrates that complexity of computing equilibrium behavior is low.

\begin{figure}[t]
\centering
\includegraphics{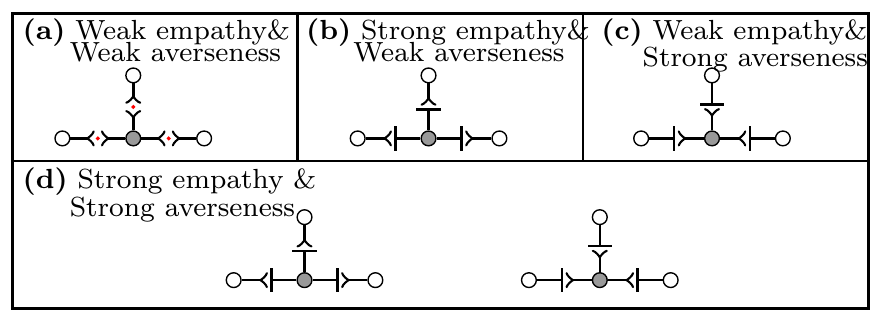} 
\caption{MMPE equilibrium strategy actions with respect to utility constants. There are $n=4$ individuals forming a star network. In cases (a)-(c) the stage equilibrium action is unique. In case (d) there two stage equilibria. When risk averseness is weak, i.e., $c_0>c_1$, all healthy individuals take action $a_i =1$ because the term inside the parantheses of the utility in \eqref{utility} is positive regardless of the action of the center individual. When empathy is weak, i.e., $c_0>3c_2$, the sick individual at the center takes action $a_i =1$ regardless of its neighbors' actions by the same reasoning. Based on these responses we can solve for stage equilibrium in (a)-(c). In (a) ($c_0>c_1$ and $c_0>3c_2$), all individuals take action 1. In (b), because all healthy individuals take action 1 due to weak averseness, the sick individual takes action 0 considering their strong empathy ($c_0<3c_2$). In (c), because the sick individual takes action 1 due to weak empathy, all healthy individuals take action 0 considering their strong averseness ($c_0<c_1$). In (d), if healthy individuals take action 1 then it is in the interest of the strongly empathetic sick individual to take action 0. However, if healthy individuals take action 0 then the sick individual receives a positive payoff from taking action 1. In both cases no individual has an incentive to deviate to another action. \label{fig:markov_game_star_example}} 
\end{figure}

We exemplify different stage Nash equilibrium actions that arise from differing utility constants ($c_0$, $c_1$ and $c_2$) on a $n=4$ individual star network with center individual sick and other individuals healthy in Fig.\ \ref{fig:markov_game_star_example}(a)-(d). We observe that in cases (a)-(c)  the equilibrium action is unique. When individuals have strong risk aversion and empathy in comparison to the socialization constant in (d), there are two stage equilibria: 1) All healthy individuals resume normal activity and the sick individual self-quarantines, and 2) sick individual resumes normal activity and healthy individuals self-quarantine. Both equilibrium strategies yield the same outcome of disease eradication -- each individual continues to take the same action until the center individual heals. However, the first equilibrium action profile yields an aggregate utility ---sum of individual payoffs--- of 3 while the second equilibrium action yields an aggregate utility of 1 yielding a ratio of $1/3$. The ratio of the worst aggregate utility value attained by an equilibrium action profile to the optimal action profile that maximizes the aggregate utility is referred to as the price of anarchy in game theory \cite{Dubey_1986}. In Appendix C, we prove a lower bound for the price of anarchy that shows it can scale with $1/n$. The set of utility constants in (d) shows that the bound is tight because the price of anarchy equals $1/(n-1) = 1/3$.

%
 
\begin{figure*}[t]
\centering
\begin{tabular}{c}
\includegraphics[width=0.8\linewidth]{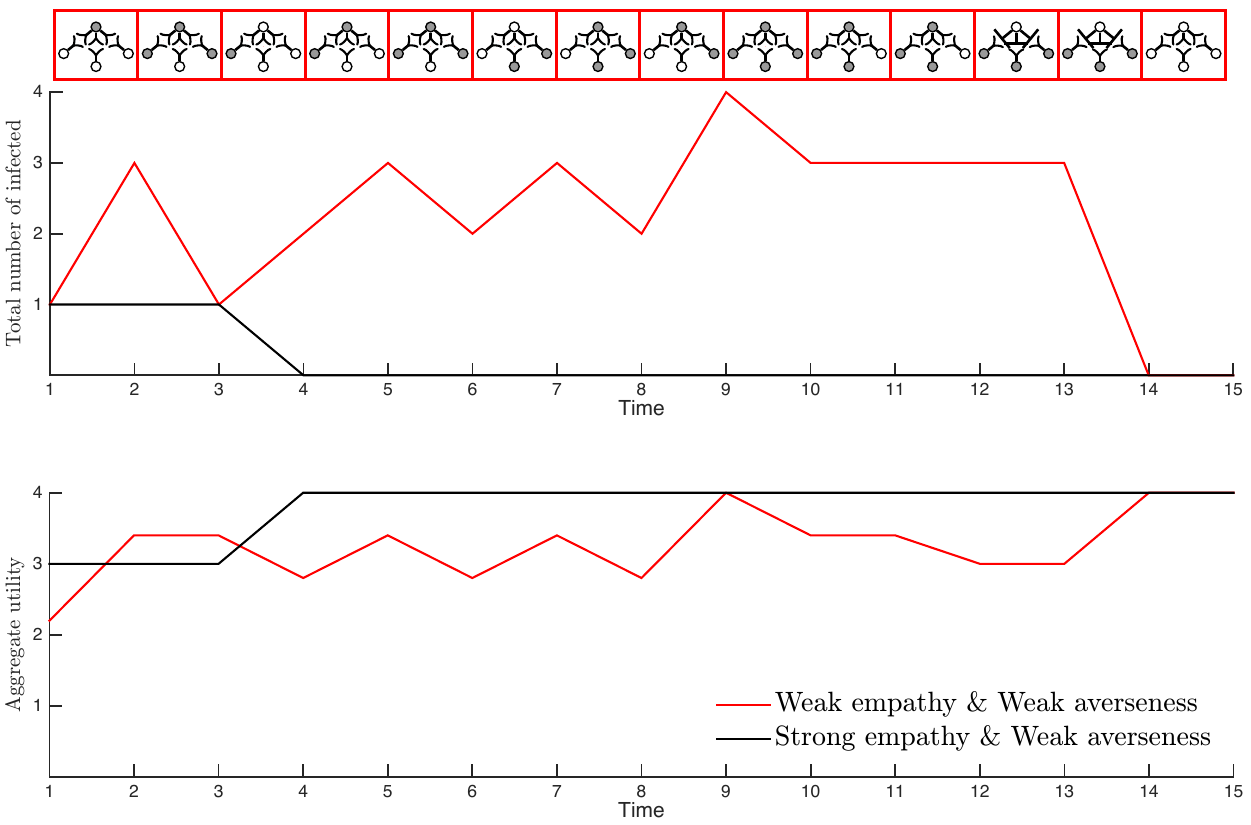} 
\end{tabular}
\caption{Behavior and disease dynamics with respect to payoff constants. The red and black lines in the figures correspond to payoff constants with weak and strong empathy, respectively. In both setups we have $\beta = 0.4$, $\delta = 0.2$, $c_0 = 1$ and weak averseness ($c_0 > c_1$). In particular, we consider risk aversion constant $c_1$ be $3c_1>c_0>2c_1$. The MMPE action for time $t\geq1$ for strong empathy is given by Fig.\ \ref{fig:markov_game_star_example}(b). The MMPE action at time $t=1$ for weak empathy is given by Fig.\ \ref{fig:markov_game_star_example}(a). The sequence of networks at the top shows the disease state and MMPE action of each individual at each time on the network for the weak empathy \& weak averseness case. In this case, the MMPE actions are such that all individuals socialize at all times unless a healthy individual has three sick neighbors when the healthy individual self-quarantines -- see times 12 and 13. This is because of the value of the risk aversion constant obeys the relation $3c_1>c_0$. Bottom figure represents the corresponding aggregate utility for each case. \label{fig:markov_game_star_demo}}
\end{figure*}

A disease is eradicated when all individuals are healthy. Note that in the Markov chain model in \eqref{eq_markov_chain} the state of all healthy individuals, $s_i(t) = 0$ for all $i \in \ccalN$ is the absorbing state. Starting at the state in Fig.\ \ref{fig:markov_game_star_example} for utility constant values given by (b)-(d), there is no chance of disease spread at an MMPE strategy profile, i.e., $p^i_{01}(t) = 0$ for all $i\in\ccalN$ and $t=1,2,\dots$. Therefore, the disease is eradicated when the center individual heals in continuation of the Markov chain dynamics. However, if constant values are as given by Fig.\ \ref{fig:markov_game_star_example}(a), there is no guarantee that the disease is eradicated when the center individual heals. 

We show the continuation dynamics of the Markov chain model starting from the cases of (a)  weak empathy \& weak averseness and (b) strong empathy \& weak averseness in Fig.\ \ref{fig:markov_game_star_demo}. In case (a), the disease takes off after the first time step in the network before eradication at time $t=14$. When its empathy is  strong in (b), disease is eradicated when the center node heals. 
We also observe that the larger aggregate utility does not always lead to a reduction in disease spread. For instance at time $t=2$ in Fig.\ \ref{fig:markov_game_star_demo}(bottom), the aggregate utility is higher in weak empathy case than the strong empathy case. In the following time $t=3$ the number of infected increases to 3 individuals for weak empathy case. At time $t=3$, the aggregate payoff remains higher than the strong empathy case with single infected individual. That is, a larger epidemic size can lead to a better aggregate utility when the empathy constant is different.

In the example given in Fig.\ \ref{fig:markov_game_star_demo}, the MMPE strategy profiles for different utility constants lead to qualitatively different outcomes. In the following sections, we explore the effects of these constants on disease spread combined with the network structure and infection parameters.   

%

\subsection{Upper bound for $R_0$ under strategic behavior}
The reproduction number $R_0$ measures the spread of an infectious disease from an initial sick individual in an otherwise susceptible host population. In the homogeneously mixed SIS model, the reproduction number is equal to $R_0 = \beta n /\delta$ \cite[Ch. 2]{Keeling_Rohani_2011}. In this model, the disease becomes an epidemic if $R_0>1$. In contact network epidemic models, $R_0>1$ is not necessarily the epidemic threshold. Rather, it is an indicator that the disease is likely to persist when there are relatively low number of infected individuals \cite{Volz_Meyers_2009}. Here, we compute an upper bound of the $R_0$ value for the stochastic SIS disease network game to relate the likelihood of disease persistence to network and utility constants.

Consider a network with degree distribution $P(k)$ where $P(k)$ denotes the probability that a selected individual has $k\in\{1,2,\dots, n-1\}$ neighbors. We assume the initial infected individual is chosen from the population uniformly at random. When individuals act according to an MMPE strategy profile, we have the following bound on $R_0$ (Appendix D),
\begin{equation} \label{r_0_bound_main}
R_0 \leq \frac{\beta}{\delta}\sum_{k=1}^{K} k P(k) \mbox{ where } K := \min(\lfloor c_0/c_2 \rfloor, n).
\end{equation}
When the empathy term $n c_2$ is smaller than $c_0$, i.e., $K=n$, we recover the bound for contact network models without behavior response. The reasoning  is as follows. An initial infected individual will continue its normal activity level ($a_i(t)=1$) even if it has $n-1$ susceptible neighbors because the expression inside the parantheses in \eqref{utility} will be positive no matter how the neighbors act. When the empathy term is larger than the socialization constant, i.e., $c_2>c_0$, the initial infected individual will not socialize ($a_i(t)=0$) even if it has a single social neighbor. Therefore, the bound above is most interesting when the empathy term is such that $c_0/n < c_2 < c_0$. In this case, only the infected individuals with connections smaller than $\lfloor c_0/c_2\lfloor$ can spread it to their neighbors initially. The degree distribution $P$ determines the frequency of individuals with number of neighbors less than $\lfloor c_0/c_2\rfloor$. We note that all the susceptible individuals will socialize at normal levels initially if $c_0 > c_1$. Still, if the disease spreads to their neighbors, their rational actions can include self-quarantine. One of the steps in the derivation of the upper bound involves assuming all susceptible neighbors of the initial infected individual remain social. Hence the risk averseness constant $c_1$ does not appear in the bound above. It is natural to think that high risk aversion may help reduce initial spread. Contrary to this intuition, when we explore the effects of $c_1$ in the following sections, we find that the risk averseness cannot stop the initial spread without a critical level of the empathy constant.  

\begin{figure} 
\centering
\begin{tabular}{c}
\includegraphics[width=0.55\linewidth]{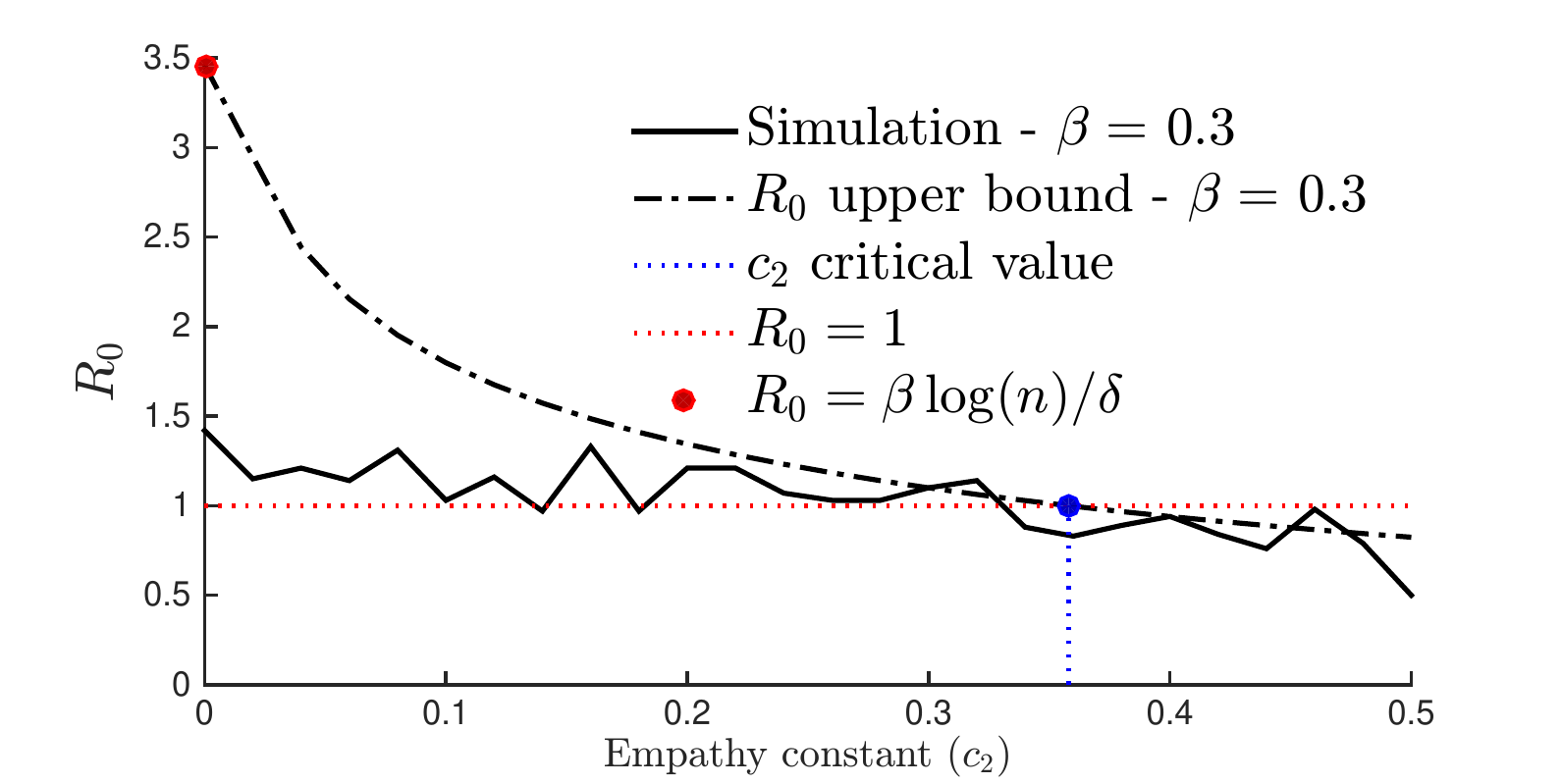} \\
\includegraphics[width=0.55\linewidth]{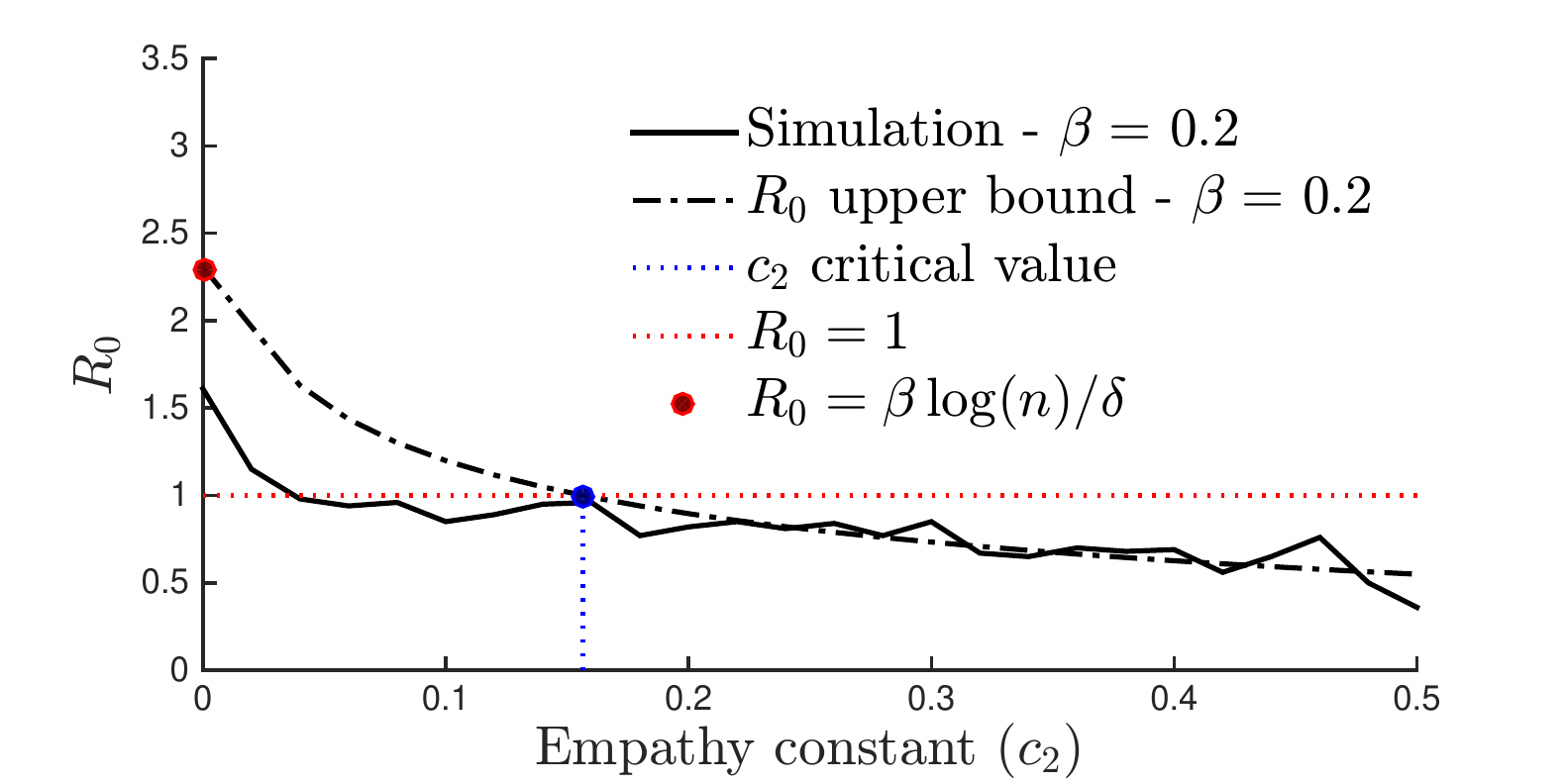} \\
\includegraphics[width=0.55\linewidth]{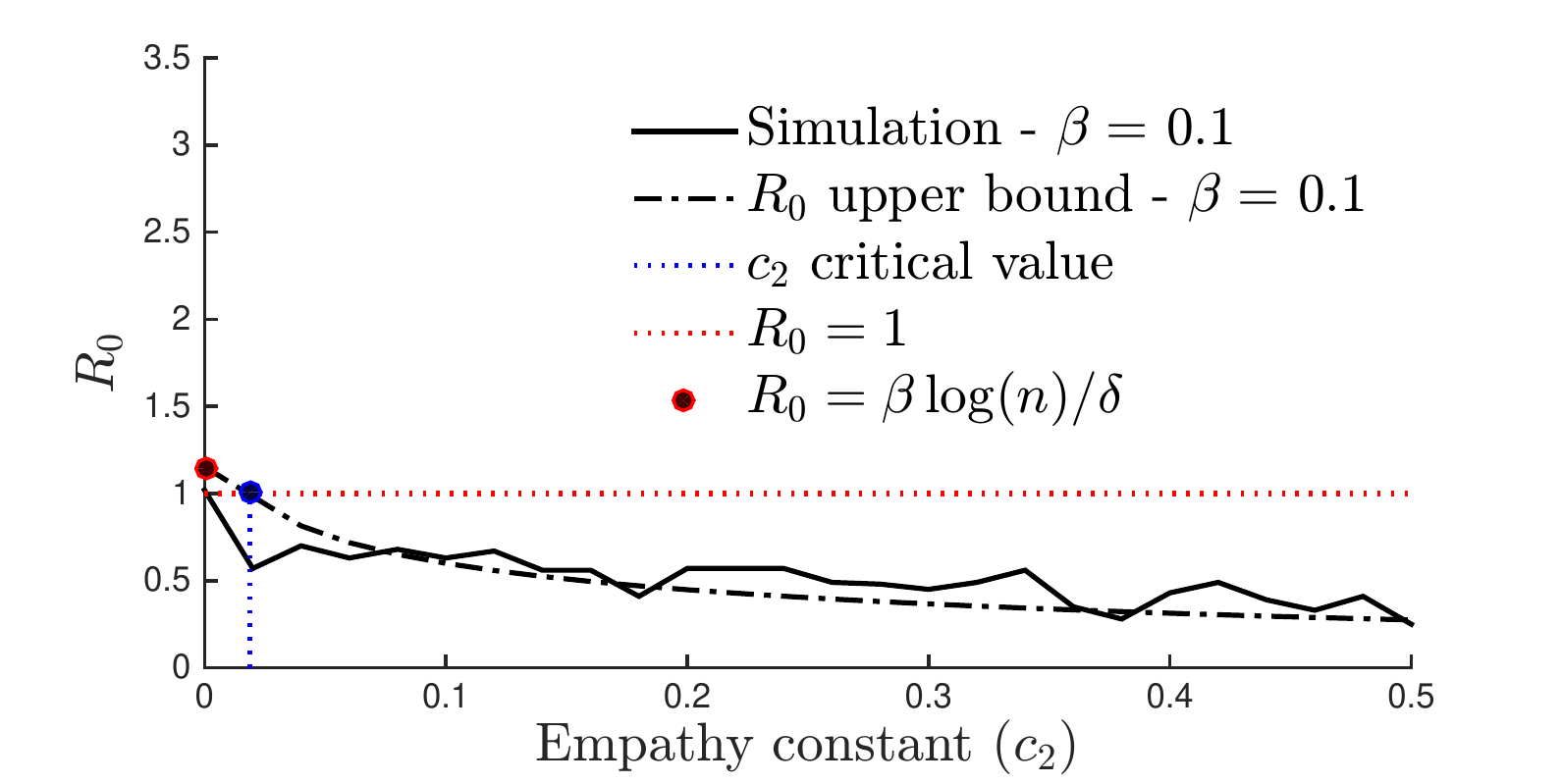} 
\end{tabular}
\caption{Upper bound accuracy for predicting critical empathy for $R_0$ threshold. We consider $n = 100$ individuals and set the constants as $\delta =0.2$, $c_0 =1$ and $c_1 = 0.24$. We let $\beta \in \{0.3, 0.2, 0.1\}$ for top, middle and bottom figures, respectively. The dotted dashed lines are the $R_0$ upper bound value \eqref{r_0_bound_scale_free_main} with respect to the $c_2$ value on $x$-axis. For $c_2 =0$, we have the red circled points corresponding to the $R_0$ upper bound when there is no behavior response by the initial sick individual, i.e., $a_i(t)=1$ for $t=1,2,\dots$. Note that all the red circled points indicate $R_0>1$. From \eqref{r_0_bound_scale_free_c_2_critical}, the critical values of $c_2$ that make $R_0<1$ equal to $0.02$, $0.16$ and $0.36$ for $\beta \in \{0.1, 0.2, 0.3\}$, respectively. These points are marked in blue. $R_0$ upper bound increases linear in $\beta$ according to \eqref{r_0_bound_scale_free_main}. We  simulate $R_0$ values as follows. We generate a scale-free network with $\gamma = 2$ according to the preferential attachment algorithm \cite{Barabasi_et_al_1999}. For each $\beta$ and $c_2$ value pair, we consider 100 realizations with randomly selecting patient zero and counting the number of individuals infected by patient zero until patient zero heals. Each point in the solid lines corresponds to the average of the total count values in 100 initializations. We observe the simulated $R_0<1$ for all $c_2$ values above the critical value in \eqref{r_0_bound_scale_free_c_2_critical}. 
\label{fig:R_0_bound}}
\end{figure}

We consider random scale free networks to illustrate the conjoint effects of network structure and the empathy constant. In a scale free network degree connectivity distribution follows power law, i.e., $P(k) \sim k^{-\gamma}$, where the value of $\gamma$ typically is in the range $[2,3]$. When $\gamma = 2$, we obtain the following upper bound for $R_0$ using \eqref{r_0_bound_main} (Appendix E),
\begin{equation} \label{r_0_bound_scale_free_main}
R_0 \leq \frac{1}{2}\frac{\beta}{\delta} \log(K+1) \mbox{ where } K := \min(\lfloor c_0/c_2 \rfloor, n).
\end{equation}
When compared with the value of  $R_0 = \beta n / \delta$ in the SIS model with homogeneous mixing, we observe how network and behavior response comes at play in the bound above. In particular if $c_0/c_2 > n$, the bound increases logarithmically with the size of the population, i.e., $\beta \log(n)/\delta$. This value is the reproduction number for the contact network SIS model with no individual behavior response to disease prevalence \cite[Ch. 17]{Newman_2010}. We observe the effect of individual behavior on $R_0$ as the empathy constant $c_2$ is increased. The power law degree distribution of the scale free network results in a logarithmic decrease in the bound with respect to increasing empathy constant $c_2$. 
From  \eqref{r_0_bound_scale_free_main} we have $R_0<1$ for 
\begin{equation}\label{r_0_bound_scale_free_c_2_critical}
c_2 > \frac{c_0}{\exp(2\delta/\beta)-1}.
\end{equation}
We measure the accuracy of the critical $c_2$ value above that makes $R_0<1$ by comparing it to simulated $R_0$ values in Fig.\ \ref{fig:R_0_bound}. Fig.\ \ref{fig:R_0_bound} shows the simulated $R_0$ values and the upper bound in \eqref{r_0_bound_scale_free_main} with respect to the $c_2$ value on the $x$-axis for a given $\beta$ value. We see that for all $\beta$ values when the upper bound is less than one, the $R_0$ value obtained from simulation is also less than 1. That is, the disease is not likely to persist when the empathy constant of individuals is above the critical level in \eqref{r_0_bound_scale_free_c_2_critical}.


\subsection{Upper bound for $R_*$ under strategic behavior}

Note that in the $R_0$ definition above the sick individual is chosen at random from the nodes. It is not difficult to see that $R_0$ can be small in a scale-free network because many individuals have very few connections while a few individuals are highly connected. Highly connected individuals are less likely to be initially infected, however, it is likely that the initial infected individual is connected to a highly connected individual \cite{Volz_Meyers_2009}. Hence, in the event that the initial individual infects its highly connected neighbor then the spread of the disease from a second infected individual can be fast. 
To account for this event, we consider the metric $R_*$ which is defined as the average number of new infections by an initial infected individual when patient zero is selected according to its connectivity degree. Consider a network with degree connectivity distribution $P(k)$. The probability that we select an infected individual with degree $k$ is given by $Q(k) := \frac{k P(k)}{\sum_k k P(k)}$. This selection process represents the likely scenario that one of the earliest infected individuals transmits the disease to a highly connected individual. We have the following upper bound for the metric given individuals acting according to an MMPE action profile (Appendix F)
\begin{equation} \label{r_x_bound}
R_* \leq \frac{\beta}{\delta}\sum_{k=1}^{K} \frac{k^2 P(k)}{\sum_{k=1}^n k P(k)}, \mbox{ where } K := \min(\lfloor c_0/c_2 \rfloor, n).
\end{equation}
Note that the denominator of the term inside the sum is the average degree connectivity of the network. The numerator is the variance of the degree distribution when $K= n$. Comparing \eqref{r_0_bound} with \eqref{r_x_bound}, it is possible that even though $R_0<1$, it might be that $R_*>1$ or vice versa. We elaborate on the differences between the two metrics further by considering the scale free network where $P(k) \sim k^{-\gamma}$ for $\gamma = 2$. We obtain the following upper bound for the $R_*$ by the inequality above (Appendix G),
\begin{equation} \label{r_x_bound_scalefree_main}
R_*\leq \frac{\beta}{\delta} \frac{K}{\log(n)}, \mbox{ where } K := \min(\lfloor c_0/c_2 \rfloor, n).
\end{equation}
Comparing the bound above with the bound for $R_0$ in \eqref{r_0_bound_scale_free_main}, we observe the $R_*$ bound is more sensitive to the empathy constant. 
In particular, the decrease in the $R_*$ bound with respect to increasing $c_2$ is linear while the  decrease is logarithmic for the $R_0$ bound. When $K=n$, the bound above grows with $n/\log(n)$ while the $R_0$ bound grows with $\log(n)$. When $K = \lfloor c_0/c_2 \rfloor$, the $R_*$ bound decreases with increasing $n$ while the bound of $R_0$ is not affected by the population size. From the bound in \eqref{r_x_bound_scalefree_main}, we have $R_*<1$ for 
\begin{equation}\label{r_x_bound_scale_free_c_2_critical}
c_2 > \frac{\beta c_0}{\delta \log(n)}. 
\end{equation}

\begin{figure} 
\centering
\begin{tabular}{c}
\includegraphics[width=.55\linewidth]{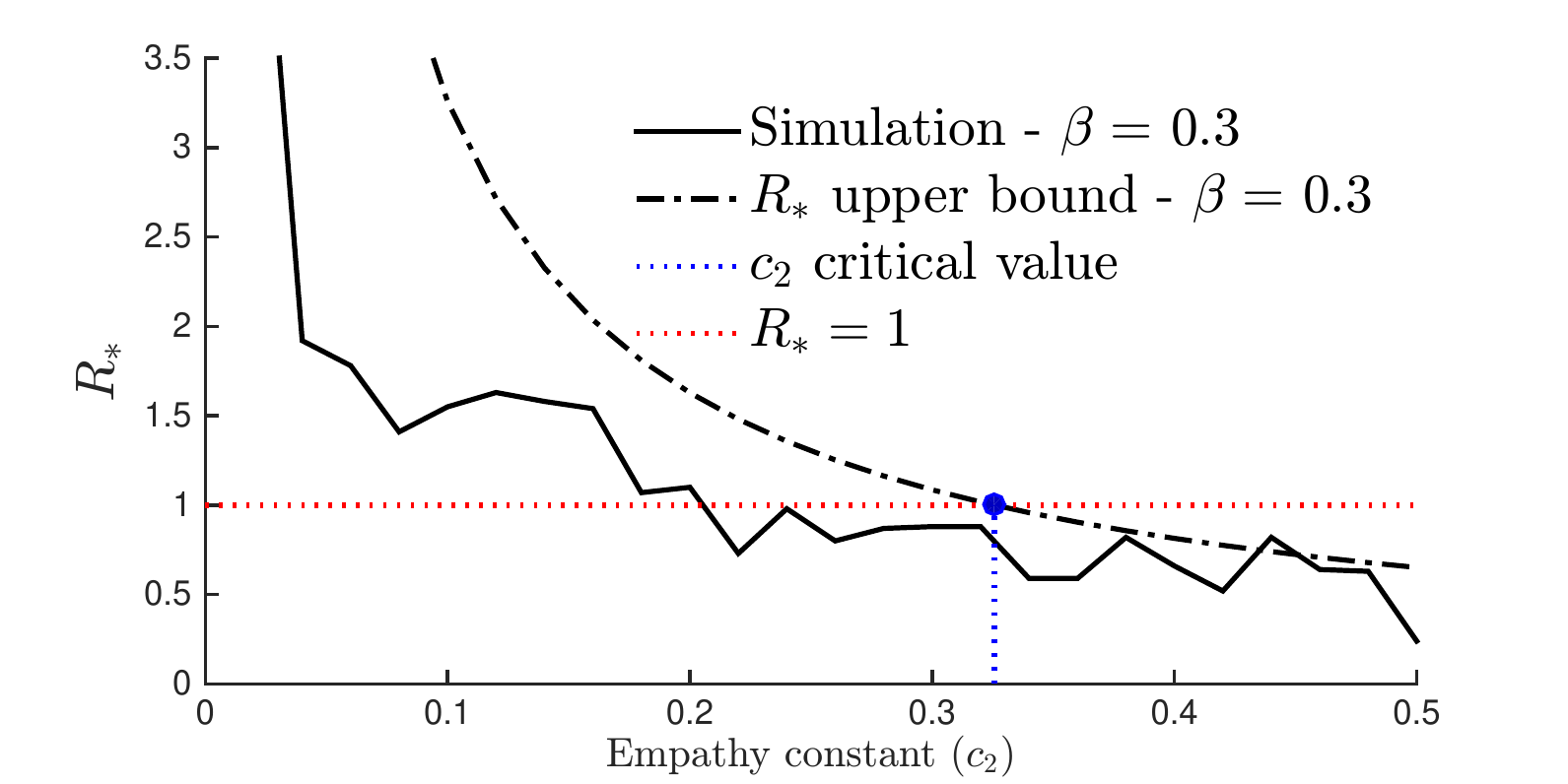} \\
\includegraphics[width=.55\linewidth]{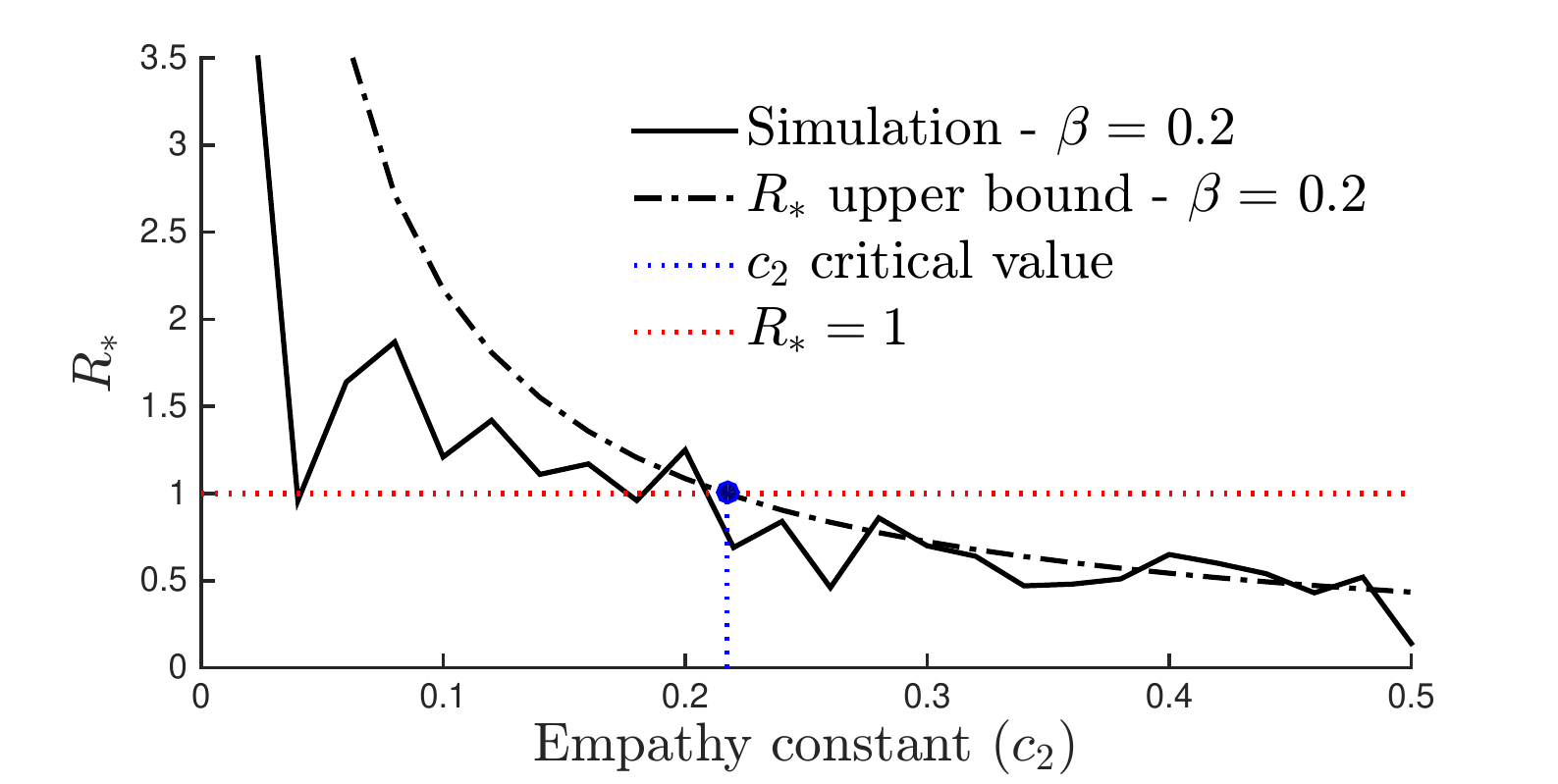} \\
\includegraphics[width=.55\linewidth]{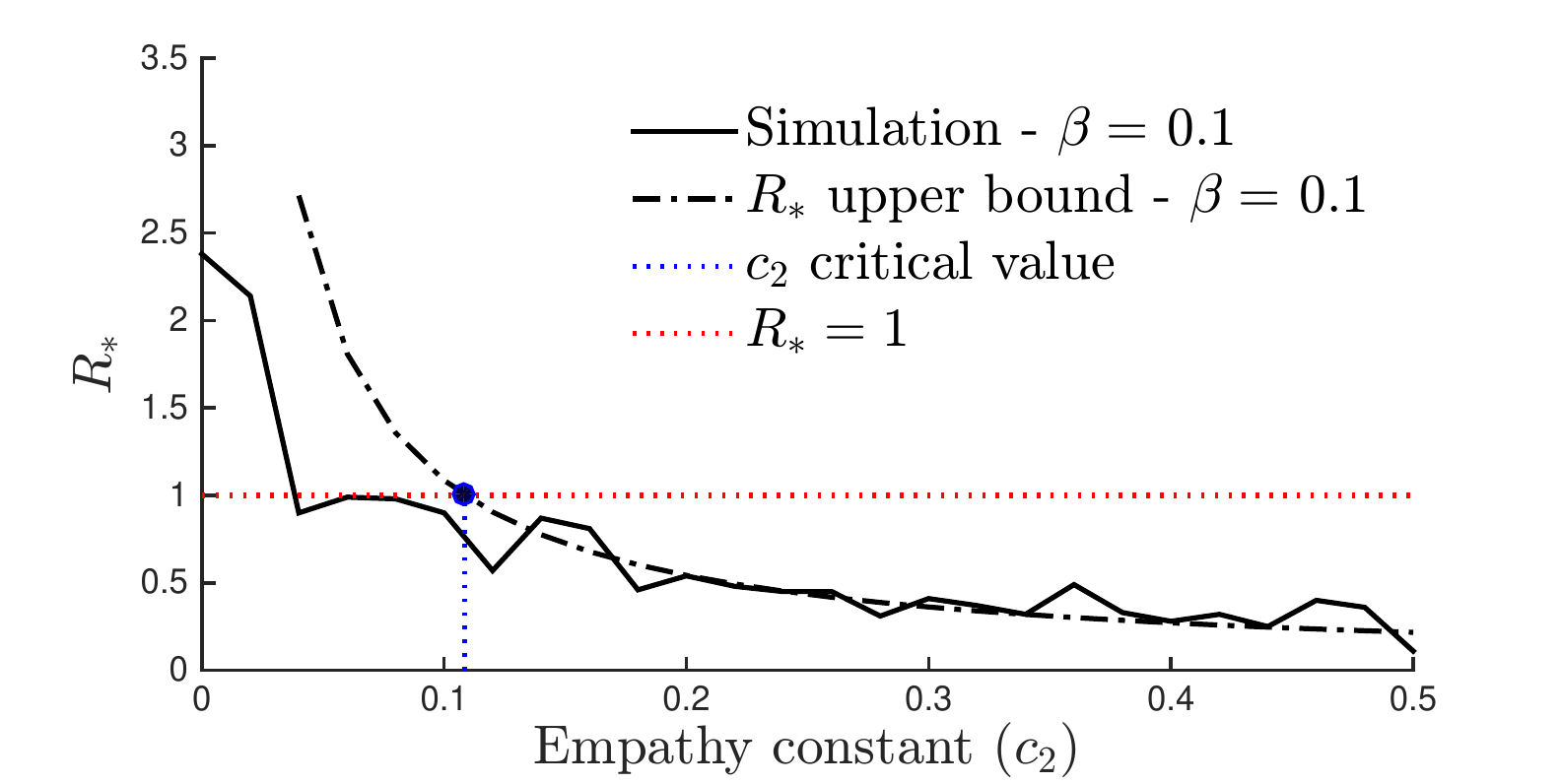} 
\end{tabular}
\caption{Upper bound accuracy for predicting critical empathy for $R_*$ threshold. We consider $n = 100$ individuals and set the constants as $\delta =0.2$, $c_0 =1$ and $c_1 = 0.24$. We let $\beta \in \{0.3, 0.2, 0.1\}$ for top, middle and bottom figures, respectively. The dotted dashed lines are the $R_*$ bound value in \eqref{r_x_bound_scalefree_main} with respect to the $c_2$ value on the $x$-axis. The critical values of $c_2$ in \eqref{r_x_bound_scale_free_c_2_critical} that make $R_*<1$ are 0.11, 0.22 and 0.33 respectively for $\beta \in \{0.1, 0.2, 0.3\}$. These points are marked in blue. We  simulate $R_*$ values identical to the way we simulate $R_0$ values in Fig.\ \ref{fig:R_0_bound} except that we select the initial sick individual according to distribution $Q(k)$. We observe the simulated $R_*$ values are less than one for all $c_2$ values above the critical value in \eqref{r_x_bound_scale_free_c_2_critical}. In comparison the critical $c_2$ values for $R_0$ in \eqref{r_0_bound_scale_free_c_2_critical} do not accurately predict the values of $c_2$ above which $R_*<1$.
\label{fig:R_0_R_x_compare}}
\end{figure}

We measure the accuracy of the critical $c_2$ value above that makes $R_*<1$ by comparing it to simulated $R_*$ values in Fig.\ \ref{fig:R_0_R_x_compare}. Fig.\ \ref{fig:R_0_R_x_compare} shows the simulated $R_*$ values where we select the initial infected individual with respect to $Q(k)$, and the upper bound in \eqref{r_x_bound_scale_free_c_2_critical} with respect to the $c_2$ value on the $x$-axis for a given $\beta$ value. We see that for all $\beta$ values when the upper bound is less than one, i.e. when $c_2$ is larger than the value in \eqref{r_x_bound_scale_free_c_2_critical}, the $R_*$ value obtained from simulation is also less than 1. That is, the disease is not likely to persist when the empathy constant of individuals is above the critical level in \eqref{r_x_bound_scale_free_c_2_critical}. In comparison, we observe that the $R_0$ upper bound \eqref{r_0_bound_scale_free_main} is not an accurate upper bound for the simulated $R_*$ values. In particular, for small values of $\beta$, while the critical $c_2$ value in \eqref{r_0_bound_scale_free_c_2_critical} estimates disease is not likely to persist, it may persist according to the $R_*$ metric.



\subsection{$R_0$ and $R_*$ as epidemic thresholds} $R_0<1$ and $R_*<1$ imply that the disease is likely to be eradicated before spreading to other individuals if initially there is a single sick individual. However, as mentioned before these conditions may not constitute a threshold for any initial state of the disease. The $R_0$ value of the SIS model with homogeneous mixing gives us a general condition for disease eradication, $\beta n/ \delta<1$. For SIS disease dynamics over networks when there is no individual response to disease prevalence, i.e., $a_i(t) = 1$  for all $i \in \ccalN$ and $t =1,2,\dots$, we obtain a disease eradication threshold of $\beta \lambda_{max}(A)/\delta<1$ where $\lambda_{max}(A)$ is the largest eigenvalue of the adjacency matrix $A$ of the contact network $G$ \cite{Mieghem_et_al_2009}.

Regarding these conditions, the more complex the model that the condition is derived from, the tighter the bound is. For instance, comparing the latter two bounds we have $\beta \lambda_{max}(A)/\delta \leq \beta n/ \delta$. That is, the networked disease model makes a sharper prediction of disease eradication than the model with homogeneous mixing assumption. By the same reasoning, the thresholds based on the stochastic network disease game constitute sharper bounds for disease eradication when compared to the networked models with no behavior response. In particular, from the bounds for $R_0$ and $R_*$ in the previous section, we know that there exist critical empathy constant values $c_2$ in \eqref{r_0_bound_scale_free_c_2_critical} and \eqref{r_x_bound_scale_free_c_2_critical} which make these values less than one even when $\beta \lambda_{max}(A)/\delta>1$. 

In Fig.\ \ref{fig_epidemic_threshold_behavior_beta}, we assess the accuracy of conditions for $R_0<1$ and $R_*<1$ as indicators of disease eradication when compared to $\beta \lambda_{max}(A)/\delta<1$. In the setup we consider disease parameter and network values where $\beta \lambda_{max}(A)/\delta> 1$, i.e.,  $\beta \lambda_{max}(A)/\delta$ value equals 2.65, 5.3, and 8 for figures left, middle and right respectively.   Furthermore, we consider two initial cases: 1) single infected (top) and 2) all infected (bottom). Fig.\ \ref{fig_epidemic_threshold_behavior_beta} shows the frequency of runs that eradicated the disease before the simulation horizon for a given set of parameter values of $\beta$, $c_1$ and $c_2$. Our initial observation is that the epidemic threshold condition $\beta \lambda_{max}(A)/\delta> 1$ is not necessarily indicative of an epidemic. That is, even though $\beta \lambda_{max}(A)/\delta> 1$,  the disease can be eradicated depending on the empathy constant value $c_2$. In addition we observe a direct relation between the frequency of disease eradication and the value of the empathy constant $c_2$. The critical values of $c_2$ that make $R_*<1$ in \eqref{r_x_bound_scale_free_c_2_critical} are indicative of disease eradication. We confirm these results for any risk averseness constant $c_1$ value in $[0,1]$. That is, the critical values of $c_2$ for which $R_*<1$ are indicators of disease eradication for all $c_1 \in [0,1]$. As $c_2$ increases above the critical value, the average time to eradication decreases (see Appendix H and I for corresponding figures). 


\subsection{The impact of risk averseness on disease spread} While the risk averseness constant $c_1$ does not show up in any of our bounds for $R_0$ and $R_*$, it can play a critical role in who takes preemptive measures as illustrated by the example in Fig.\ \ref{fig:markov_game_star_example}. Thus, the equilibrium infectivity level can depend on $c_1$. Our first result regarding the risk averseness constant shows that when the empathy constant $c_2$ is zero, we obtain the same epidemic threshold condition as when the behavior response is not accounted for in a disease spread model, i.e., $\beta \lambda_{max}(A)/\delta>1$ for any $c_1 \in [0,1]$. This threshold is analytically obtained by first approximating the Markov chain dynamics by a $n$-state differential equation and then linearizing the approximate model around its trivial fixed point, the origin -- see Appendix J. The derivation is similar to the derivation of the threshold for disease dynamics over networks without behavior response \cite{Mieghem_et_al_2009}. This result implies that no matter how risk averse the susceptible individuals are, they cannot eradicate the disease with certainty without the empathy of infected individuals in the stochastic disease network game. 

\begin{figure*} 
\centering
\includegraphics[width=0.95\linewidth]{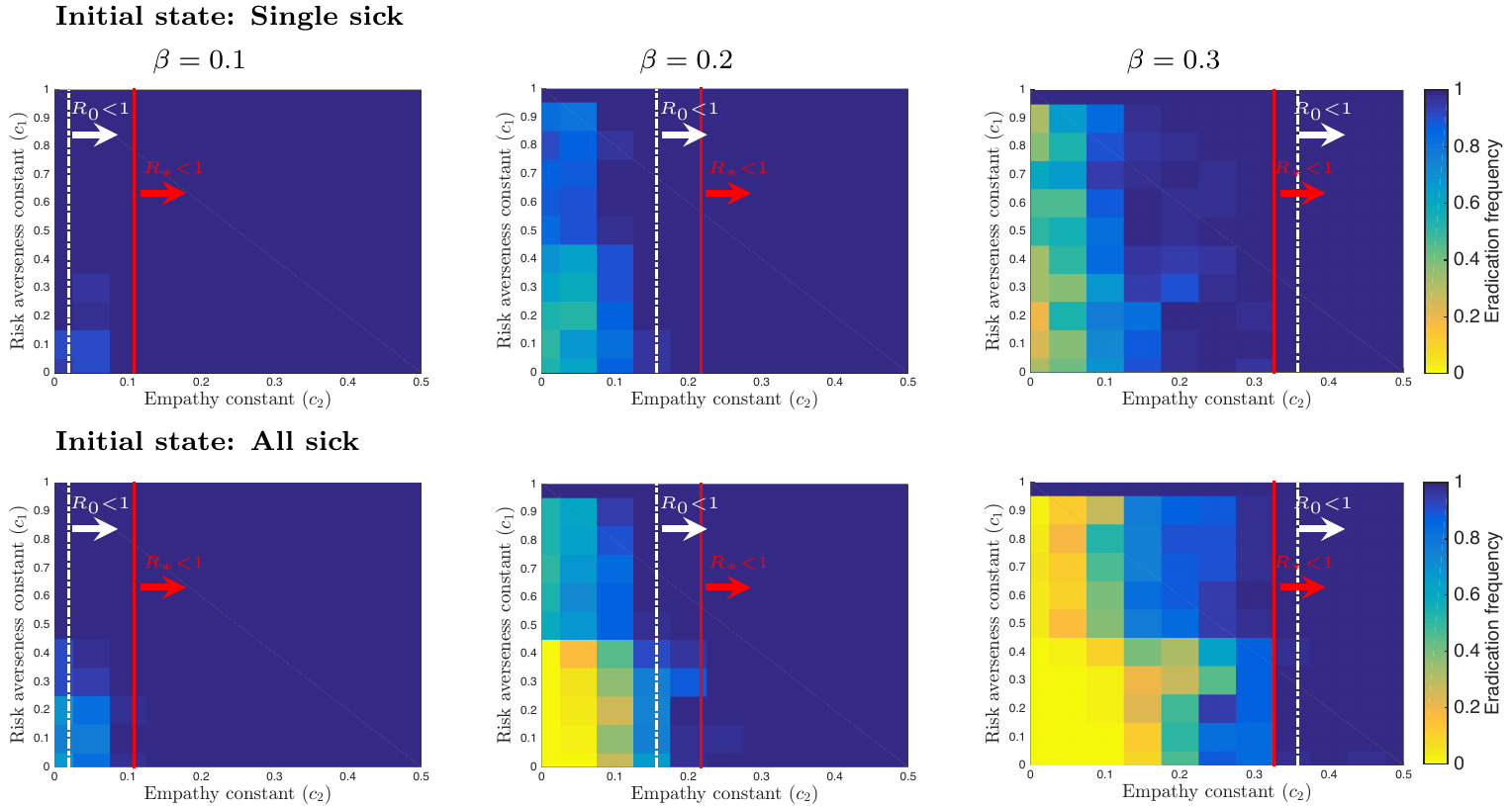} 
\caption{Effect of risk averseness $c_1$ and empathy $c_2$ constants on eradication. We let $n=100$, $\delta = 0.2$, and $c_0 = 1$. The infection rate $\beta$ values equal to 0.1,0.2 and 0.3 for figures left, middle and right, respectively. At the top the runs start with single infected individual, and at the bottom all individuals infected. The axes in each plot correspond to the constant values of $c_1$ and $c_2$. For a given value of $c_1$ and $c_2$, we generate 50 scale-free networks using the preferential attachment algorithm and run the stochastic disease network game for $200$ steps for each network. The grid color represents the ratio of runs in which disease is eradicated within $200$ steps. For figures left, middle, and right the eradication of threshold $\beta \lambda_{max}(A)/\delta$ is equal to 2.65, 5.3, and 8, respectively. That is, $\beta \lambda_{max}(A)/\delta>1$ for all the figures. The critical values of $c_2$ $\{0.02, 0.16, 0.36\}$ that make $R_0<1$ for $\beta = \{0.1, 0.2, 0.3\}$ calculated using \eqref{r_x_bound_scale_free_c_2_critical} are marked with white dotted dashed lines. The critical values of $c_2$ $\{0.11, 0.22, 0.33\}$ that make $R_*<1$ for $\beta = \{0.1, 0.2, 0.3\}$ calculated using \eqref{r_x_bound_scale_free_c_2_critical} are accurate in determining fast eradication for any value of $c_1$ (marked with red solid lines).
}
\label{fig_epidemic_threshold_behavior_beta}
\end{figure*}

Both this analytical result and the bounds for $R_0$ and $R_*$ assume initially only a single individual is infected. Hence, these conditions might not be accurate when the infected individuals are large. In Fig.\ \ref{fig_epidemic_threshold_behavior_beta} (bottom) we consider numerical simulations where the whole population is infected initially. The numerical simulations confirm that the disease cannot be eradicated at any $c_1$ value if $\beta \lambda_{max}(A)/\delta$ is large and the empathy constant is zero. For values of $\beta \lambda_{max}(A)/\delta$ closer to 1, a high enough risk aversion helps to eliminate the disease Fig.\ \ref{fig_epidemic_threshold_behavior_beta} (left). Finally, we observe that the frequency of disease eradication increases as we increase the $c_1$ value when $c_2$ is positive in Fig.\ \ref{fig_epidemic_threshold_behavior_beta} (middle and right). 

This observation implies that even a little bit of empathy can go a long way in eradication of the disease given risk averse susceptible individuals. In other words, when the empathy term is a positive value $c_2>0$, there is high enough risk averseness constant that is likely to eradicate the disease. While it may be that the risk averseness by itself cannot eradicate the disease, we observe that it reduces the average infectivity level when the disease is endemic  (see Appendix I for corresponding figures). 


\section{Discussion}

Behavior changes are ubiquitous during infectious disease outbreaks. Hence, their accurate modeling can help with the prediction of disease's impact and with the assessment of policy measures. To this end, we considered a stochastic network game where individuals respond to the current risk of disease spread, and their responses together with the current state of the disease and the contact network structure stochastically determine the next stage of the disease. In particular the game is played among the healthy and the sick in an SIS infectious disease. In our scenario, the concern for disease contraction of a healthy individual increased with the number of sick contacts that are not taking any preemptive measures. Similarly, sick individuals had increased concerns for disease spread when there are more healthy contacts that do not take protective measures. This meant that the incentives for a healthy individual taking a measure decreased as more of its sick contacts took preemptive measures, e.g., staying at home. Similarly, the incentive for preemptive measures decreased for sick individuals as the healthy got more cautious. The consequences of these incentives are not trivial in a disease contact network setting where an individual cares about the behaviors of its neighbors who themselves care about their neighbors and so on. Hence, our analysis focused on the impact of rational behavior on disease spread.

Our results showed that when individuals act rationally, there exists a level of concern by the infected individuals (empathy) above which the reproduction number is less than one. We have confirmed this result for two notions of the reproduction number. In the first notion ($R_0$), an initial infected individual is selected uniformly at random from the population. For this notion, we showed that the critical empathy term increases exponentially with the inherent infection rate of the disease for a scale-free network. In the second notion of the reproduction number ($R_*$), the initial infected individual is selected randomly, with probability proportional to its number of contacts. For this notion, we showed the critical empathy term increases linearly with the inherent infection rate of the disease for a scale-free network. We found that these critical levels of empathy for $R_0$ and $R_*$ are also good indicators of disease eradication starting from any initial disease state. In contrast, the risk aversion of healthy individuals is not a determinant in disease eradication when the empathy of the sick individuals is zero, i.e., when sick individuals are not responsive. Yet, for a positive level of empathy, there exists a risk aversion constant above which disease is likely to be eradicated. 

Previous behavior response models to disease spread considered protective measures, e.g., social distancing, predominantly by the susceptible individuals \cite{Wang_et_al_2015}. Here we incorporated the response of the infected individuals as individuals willing to reduce their risk of transmitting to others. Our results show that their behavior is more critical than the risk aversion of susceptible individuals. This imbalance of roles played by the response of the infected versus the susceptible individuals in disease eradication affords critical policy insights. In particular, it suggests that public health recommendations should emphasize the common practices when sick, e.g., covering cough, not going to work. It also highlights the importance of accounting for individual response in predictions of disease spread. It is worth noting that we can interpret the behavior response of infected individuals as altruism as termed and noted empirically \cite{Shim_et_al_2012,Steelfisher_et_al_2010} since their worry is to refrain from infecting their neighbors. We also remark that in an SIS infectious disease model it is also in the self-interest of an infected individual  to not spread the disease as the longer the disease prevails the higher the chance that the individual will contract the disease again.

This study focused on explaining the effects of individual measures during an epidemic. In reality, these individual measures are coupled with policies implemented by public health institutions. As a future research direction, it would be interesting to couple the decentralized individual reactions to disease prevalence with centralized policies such as public information, quarantining individuals that are sick or vaccination campaigns. Another potential direction could be to modify the benchmark assumption that individuals behave rationally. Analysis of the sensitivity of the results to deviations from the rationality is of interest but it is not trivial how these behavioral deviations would be modeled \cite{Funk_et_al_2015}. Finally, the present model considered an SIS disease spread. Despite its simplifications, the current model provides a principled approach to connect rational decision making with other complex dynamic disease models. We are hopeful that further extensions will provide insights on how to influence the short- and long-term behavior of individuals so as to reduce the spread and burden of infectious disease.

\appendix

\section{Utility lower bound} \label{utility_lower_bound}
First, we derive the following upper bound on $p_{01}^i$, 
\begin{align}
p_{01}^i = 1-\prod_{j\in \ccalN_i} \left(1 - \beta a_i a_j s_j(t)\right) \leq \beta a_i \sum_{j \in \ccalN_i} a_j s_j(t). \label{eq:upper_bound_transition}
\end{align}
We replace $p_{01}^i$ in the preference function with the upper bound above to get the utility function 
\begin{align} \label{utility}
 u_i(a_i, \{a_{j}\}_{j\in \ccalN_i}, s(t)) = 
  a_i \beta \bigg({c_0}  -  c_1 (1-s_i(t)) \sum_{j\in \ccalN_i} a_j s_j(t)  - { c_2 s_i(t) \sum_{j\in\ccalN_i} a_j (1-s_j(t))}\bigg),
\end{align}
where we redefined $c_0 = \beta c_0$. This is a lower bound for the preference function in equation [2] of the manuscript.

\section{Existence of an MMPE strategy profile} \label{existence_proof}

Define the space of probability distributions on the action space ($0\leq a_i \leq 1$) $[0,1]$ as $\bigtriangleup([0,1])$. A mixed strategy profile $\sigma(\cdot)$ is a function that maps the state $s \in \{0,1\}^n$ to the space of probability distributions on the actions space, i.e., $\sigma_i:  \{0,1\}^n \to \bigtriangleup([0,1])$. The definition of a mixed  MMPE strategy profile $\sigma^* := \{\sigma_i^*:\{0,1\}^n \to \bigtriangleup[0,1]\}$ is a distribution on the action space that satisfies the following, for all $t=1,2,\dots$, and $i \in \ccalN$,
\begin{align} \label{Nash_definition}
u_i(\sigma_i^{*}(s(t)), \sigma_{\ccalN_i}^{*}(s(t)), s(t)) \geq  u_i(\sigma_i(s(t)), \sigma_{\ccalN_i}^{*}(s(t)), s(t)) 
\end{align}
for any $\sigma_i \in \{0,1\}^n \to \bigtriangleup([0,1])$ where $\sigma_{\ccalN_i}^{*} := \{\sigma_{j}^{*}: j\in \ccalN_i \}$. From the existence of a mixed Nash equilibrium in games with continuous payoffs, we know that a mixed MMPE exists for the game with payoffs \eqref{utility}. 

In the following we constructively show that in the stochastic disease network game with payoffs in \eqref{utility}, a degenerate (pure) MMPE strategy profile $\sigma^* := \{\sigma_i^*:\{0,1\}^n \to [0,1]\}$ that satisfies the above relation in \eqref{Nash_definition} exists.  Note that a degenerate distribution puts weight one on a single action value, that is, the strategy profile corresponds to a single action profile for any state. 

Since in the stochastic game population response is determined by the current state of the disease only, i.e., equilibrium is stationary, it suffices to show the existence of a pure Nash equilibrium strategy for the stage game with state $s \in \{0,1\}^n$. A pure Nash equilibrium (NE) strategy profile $\sigma^*:\{0,1\}^n \to [0,1]^n$ of the stage game with payoffs \eqref{utility} and state $s$ satisfies 
\begin{align} \label{Nash_definition_stage}
u_i(\sigma_i^{*}, \sigma_{\ccalN_i}^{*}, s) \geq  u_i(\sigma_i, \sigma_{\ccalN_i}^{*}, s) 
\end{align}
for any $\sigma_i \in \{0,1\}^n \to \bigtriangleup([0,1])$. We define the corresponding equilibrium action profile as $a^* := \sigma^*(s)$ for a given state $s$. 
%

For a given individual $i$ with its state $s_i$, its neighbors' state $s_{\ccalN_i}:= \{s_j\}_{j \in \ccalN_i}$ and neighbor action profile $a_{\ccalN_i}$ we have the best response of individual $i$ as follows,
\begin{align}
BR_i(a_{\ccalN_i}, s_i, s_{\ccalN_i}) &:= \argmax_{a_i \in [0,1]} u_i(a_i, a_{\ccalN_i}, s_{\ccalN_i}) \label{best_response} \\
& = \bbone\left({c_0} >  c_1 (1-s_i) \sum_{j\in \ccalN_i} a_j s_j + { c_2 s_i \sum_{j\in\ccalN_i} a_j (1-s_j)}\right) \label{best_response_specific}
\end{align}
where $\bbone(\cdot)$ is the indicator function. Since the payoffs are linear in self-actions, the actions that maximize the payoffs are in the extremes -- $a_i =1$ or $a_i = 0$ -- depending on the states and actions of their neighbors. 
We can equivalently represent the NE definition in \eqref{Nash_definition_stage} as a fixed point equation by using the best response definition,  
\begin{align} \label{NE_best_response}
a_i^{*} = BR_i(a_{\ccalN_i}^{*}, s_i, s_{\ccalN_i}) \quad \forall i\in \ccalN.
\end{align}
In the following we define the concept of strictly dominated action which will be a useful in finding the Nash equilibrium. 
\begin{definition}[Strictly dominated action] \label{strictly_dominated_definition}
For a given state $s\in \{0,1\}^n$, an action $a_i \in [0,1]$ is strictly dominated if and only if there exists an action $a_i' \in [0,1]$ such that 
\begin{equation}\label{eq:strictly_dominated_definition}
u_i(a_i', a_{\ccalN_i}, s_i, s_{\ccalN_i}) >  u_i(a_i, a_{\ccalN_i}, s_i, s_{\ccalN_i}) \quad \forall a_{\ccalN_i}
\end{equation}
\end{definition}
If an action $a_i$ is strictly dominated then there exists a more preferable action $a_i'$ for any circumstance. It is clear that if an action is strictly dominated then it cannot be a rational action from \eqref{NE_best_response}. In a game we can iteratively remove the strictly dominated actions, this process is called the iterated elimination of strictly dominated strategies and is defined below.
\begin{definition}[Iterated elimination of strictly dominated actions] \label{def:elimination_strict}
Set the initial set of actions $A_i^0 = [0,1]$ for all $i$, and for any $k\in \naturals$ set 
\begin{equation}
A_i^k = \{a_i \in A_i^{k-1} \given a_i \text{ is not strictly dominated given any } a_{\ccalN_i} \in A_{\ccalN_i}^{k-1}\}
\end{equation}
\end{definition}
We denote the set of actions of individual $i$ that survive the iterated elimination by $A_i^\infty := \bigcap_{k =0}^\infty A_i^k$. When $A_i^\infty$ has a single element, we say $A_i^\infty$ is a singleton. 

\begin{figure}[t]
\centering
\includegraphics{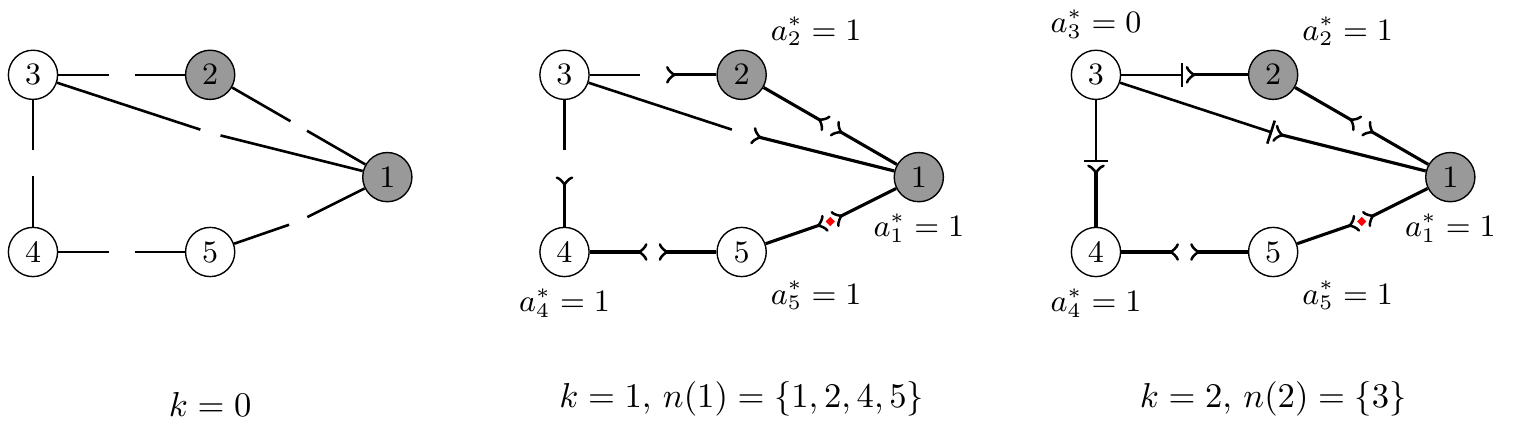} 
	\caption{\small The iterated elimination process in Definition \ref{iterated_elimination_process} on a 5 individual network. The payoff constants are such that $2 c_1 > c_0> c_1$ and {$c_0 > 2 c_2$}. Individuals 1 and 2 are sick ($s_1=1$, $s_2 = 1$) and the rest are healthy. The contact network is shown at $k=0$. At step $k=1$, individuals in $n(1) = \{1,2,4,5\}$ eliminate all actions except 1 by \eqref{odd_case_response} given $c_0 > 2 c_2$. Individual 3 cannot eliminate any action from the space. At step $k=2$, individual 3 eliminates all actions except 0 by \eqref{even_case_response}. Since all individuals have singleton non-dominated action spaces, i.e., $\ccalN \setminus (n(1) \bigcup n(2)) = \emptyset$, the process ends. Furthermore, the corresponding action profile is a Nash equilibrium of the stage game in \eqref{utility} by Lemma \ref{NE_action_profile}. 
	  }
	\label{fig:algorithm_example}
\end{figure}

We formally state an iterated elimination process for the game in \eqref{utility}. 

\begin{definition}[An iterated elimination process for the epidemic game] \label{iterated_elimination_process}
Begin with action spaces $A_i^0 = [0,1]$ for all $i \in \ccalN$. Fix $s \in \{0,1\}^n$. Iterate $k = 1,2,\dots$
\begin{enumerate}
\item[1)] [$k=odd$] Let $A_i^k = 1$ for all $i \in n(k)$ where 
\begin{equation}
n(k) := \left\{i\in\ccalN\setminus \bigcup_{l=1}^{k-1} n(l) : {c_0} >  \left(c_1 (1-s_i) \hspace{-12pt}\sum_{j\in \ccalN_i \setminus \bigcup_{l=even} n(l)}  \hspace{-12pt} s_j + {c_2 s_i  \hspace{-12pt}\sum_{j\in\ccalN_i \setminus \bigcup_{l=even} n(l)}  \hspace{-12pt}(1-s_j)}\right) \right\} \label{odd_case_response}
\end{equation}
\item[2)] [$k=even$] Let $A_i^k = 0$ for all $i \in n(k)$ where 
\begin{equation}
n(k) := \left\{i\in\ccalN\setminus \bigcup_{l=1}^{k-1} n(l) : {c_0} <  \left(c_1 (1-s_i) \hspace{-12pt}\sum_{j\in \ccalN_i \bigcap \bigcup_{l=odd} n(l)}  \hspace{-12pt} s_j + {c_2 s_i  \hspace{-12pt}\sum_{j\in\ccalN_i \bigcap \bigcup_{l=odd} n(l)}  \hspace{-12pt}(1-s_j)}\right) \right\} \label{even_case_response}
\end{equation}
\end{enumerate}
If $n(k) = \emptyset$ then stop iteration of $k$.
\end{definition}
Our next result shows that the process described above eliminates all strictly dominated strategies in finite time. 
\begin{lemma} \label{iterated_elimination_lemma}
Consider the process described in Definition \ref{iterated_elimination_process}. There exists an iteration step $k \leq n$ such that $\{A_i^k\}_{i \in \ccalN}$ are the set of actions that are not strictly dominated for the game with payoffs \eqref{utility} and state $s \in \{0,1\}^n$.  
\end{lemma}
\begin{proof}
First consider the odd iteration steps, $k=odd$. If the condition inside the bracket in \eqref{odd_case_response} holds then for any action that individual $j\in \ccalN_i \setminus \bigcup_{l=even} n(l)$ takes, $a_i'=1$ dominates any other action $a_i\in [0,1]\setminus \{1\}$ by Definition \ref{strictly_dominated_definition}. To see this first recall the best response of individual $i$ \eqref{best_response_specific}. Note that by the even steps, individuals $j\in  \bigcup_{l=even} n(l)$ have $a_j=0$ as the only not dominated action. Hence the best response of individual $i$ can be rewritten as 
\begin{equation}
\bbone\left({c_0} >  \left(c_1 (1-s_i) \hspace{-12pt}\sum_{j\in \ccalN_i \setminus \bigcup_{l=even} n(l)}  \hspace{-12pt} a_j s_j + {c_2 s_i  \hspace{-12pt}\sum_{j\in\ccalN_i \setminus \bigcup_{l=even} n(l)}  \hspace{-12pt}a_j (1-s_j)}\right)\right)
\end{equation}
where we remove neighbors that only have zero as the not dominated action. If the inequality inside the indicator function is true even when the remaining neighbors of $i$ take action $1$, then $a_i=1$ dominates all the other actions $[0,1)$ of $i$, that is, for all $a_j \in [0,1]$ for $j \in \ccalN \setminus i$,
\begin{align} \label{eq:socialize_in_the_worst_case}
 \bbone&\left({c_0} >  \left(c_1 (1-s_i) \hspace{-12pt}\sum_{j\in \ccalN_i \setminus \bigcup_{l=even} n(l)}  \hspace{-12pt} s_j + {c_2 s_i  \hspace{-12pt}\sum_{j\in\ccalN_i \setminus \bigcup_{l=even} n(l)}  \hspace{-12pt} (1-s_j)}\right)\right) \leq \nonumber\\
 &\bbone\left({c_0} >  \left(c_1 (1-s_i) \hspace{-12pt}\sum_{j\in \ccalN_i \setminus \bigcup_{l=even} n(l)}  \hspace{-12pt} a_j s_j + {c_2 s_i  \hspace{-12pt}\sum_{j\in\ccalN_i \setminus \bigcup_{l=even} n(l)}  \hspace{-12pt}a_j (1-s_j)}\right)\right) 
\end{align}
Assuming the left hand side of the above inequality is one then by the definition of strictly dominated action in Definition \ref{strictly_dominated_definition}, $a_i =1$ is the only action that is not dominated. 

Next, we consider the even iteration steps, $k=even$. If the condition inside the bracket in \eqref{even_case_response} holds then for any action that individual $j\in \ccalN_i \setminus \bigcup_{l=odd} n(l)$ takes, $a_i'=0$ dominates $a_i\in [0,1]\setminus \{0\}$ by Definition \ref{strictly_dominated_definition}. To see this first recall the best response of individual $i$ \eqref{best_response_specific}. Note that by the odd steps, individuals $j\in  \bigcup_{l=odd} n(l)$ have $a_j=1$ as the only not dominated action. Hence the best response of individual $i$ can be rewritten as 
\begin{align}
\bbone \Bigg(c_0 &>  \Bigg(c_1 (1-s_i) \hspace{-2pt}\bigg(\sum_{j\in \ccalN_i \setminus \bigcup_{l=odd} n(l)}  \hspace{-2pt} a_j s_j +\sum_{j\in \ccalN_i \bigcap \bigcup_{l=odd} n(l)}  \hspace{-2pt} s_j \bigg)\nonumber \\
&+ c_2 s_i  \hspace{-2pt}\bigg(\sum_{j\in\ccalN_i \setminus \bigcup_{l=odd} n(l)}  \hspace{-2pt}a_j (1-s_j) + \sum_{j\in\ccalN_i \bigcap \bigcup_{l=odd} n(l)}  \hspace{-2pt} (1-s_j)\bigg)\Bigg)\Bigg)
\end{align}
where we separate neighbors $j\in\ccalN_i \bigcap \bigcup_{l=odd} n(l)$ that only have $a_j = 1$ as the not dominated action from the other neighbors $j\in\ccalN_i \setminus \bigcup_{l=odd} n(l)$. If the inequality inside the indicator function is false even when the remaining neighbors of $i$ take action $0$, then $a_i=0$ dominates all the other actions $(0,1]$ of $i$, that is, for any $a_j \in [0,1]$ for $j\in\ccalN_i \setminus \bigcup_{l=odd} n(l)$
\begin{align} \label{eq:not_socialize_in_the_best_case}
\bbone \Bigg(c_0 <  \Bigg(c_1 &(1-s_i) \hspace{-2pt}\bigg(\sum_{j\in \ccalN_i \bigcap \bigcup_{l=odd} n(l)}  \hspace{-2pt} s_j \bigg)
+ c_2 s_i  \hspace{-2pt}\bigg(\sum_{j\in\ccalN_i \bigcap \bigcup_{l=odd} n(l)}  \hspace{-2pt} (1-s_j)\bigg)\Bigg)\Bigg)
 \leq \nonumber\\
\bbone \Bigg(c_0 <&  \Bigg(c_1 (1-s_i) \hspace{-2pt}\bigg(\sum_{j\in \ccalN_i \setminus \bigcup_{l=odd} n(l)}  \hspace{-2pt} a_j s_j +\sum_{j\in \ccalN_i \bigcap \bigcup_{l=odd} n(l)}  \hspace{-2pt} s_j \bigg)\nonumber \\
&+ c_2 s_i  \hspace{-2pt}\bigg(\sum_{j\in\ccalN_i \setminus \bigcup_{l=odd} n(l)}  \hspace{-2pt}a_j (1-s_j) + \sum_{j\in\ccalN_i \bigcap \bigcup_{l=odd} n(l)}  \hspace{-2pt} (1-s_j)\bigg)\Bigg)\Bigg)
\end{align} 
Assuming the left hand side of the above inequality is one then by the definition of strictly dominated action in Definition \ref{strictly_dominated_definition}, $a_i =0$ is the only action that is not dominated. 

Further if no individual can eliminate a dominated strategy $n(k) = \emptyset$ at an iteration $k$ then at following iterations $k+1, k+2, \dots$ there won't be any individuals that eliminate any actions as strictly dominated. To see this, assume the opposite is true, that is, $n(k+1) \neq \emptyset$. Since $n(k)=0$ then the condition in $n(k+1)$ is identical to the conditions inside \eqref{odd_case_response} or \eqref{even_case_response} for $n(k-1)$ depending on whether $k+1$ is odd or even, respectively. There cannot be an individual that satisfies the conditions at iteration $k+1$ because $n(k+1)$ is selected among individuals that have not been previously in any set, i.e., $i \in \ccalN\setminus \bigcup_{l=1}^{k} n(l)$. This contradicts $n(k+1) \neq \emptyset$. As a result, at each iteration $k$ the number of individuals in $n(k)$ has to be positive until an iteration step $n(k)$ at which either there is no individual left $\ccalN\setminus \bigcup_{l=1}^{k} n(l) = \emptyset$ or there is no individual that satisfies the condition after the colon in \eqref{odd_case_response} or \eqref{even_case_response}. 

Suppose now that the process stops at iteration $k$, i.e., $n(k) = \emptyset$, but there exists an individual $i \in \ccalN\setminus \bigcup_{l=1}^{k-1} n(l)$ with strictly dominated action \eqref{eq:strictly_dominated_definition} $a_i \in A_i^k$. Suppose $k$ is odd then by \eqref{eq:socialize_in_the_worst_case}, the action $a_i \in A_i^k$ must be dominated by $a_i' = 1$. Suppose $k$ is even, then by \eqref{eq:not_socialize_in_the_best_case} the action $a_i \in A_i^k$ must be dominated by $a_i' = 0$. Then $n(k) \neq \emptyset$ which is a contradiction. This together with the fact that if $n(k) = \emptyset$ then $n(l)=\emptyset$ for $l>k$ implies that if the process in Definition \ref{iterated_elimination_process} stops all the strictly dominated strategies are eliminated. 

Finally, the fact that the number of elements of $n(k)$ is positive at every iteration except the last iteration in Definition \ref{iterated_elimination_process} implies that the iteration ends in at most $n$ iterations.
\end{proof}

The following Lemma shows that once all the strictly dominated strategies are eliminated, the action profile that assigns socialize to susceptible individuals and do not socialize to infected individuals is a pure Nash equilibrium strategy profile.  
\begin{lemma} \label{NE_action_profile}
Consider the game defined by the payoffs in \eqref{utility} given state $s \in \{0,1\}^n$ and the iterated elimination process in Definition \ref{iterated_elimination_process}. Denote the action space of $i\in \ccalN$ that is not strictly dominated by $A_i^*$. Let $a_i^{*} = A_i^*$ if $A_i^*$ is a singleton. If $A_i^*$ is not a singleton then let $a_i^{*} =1$ for $s_i = 0$ and $a_i^{*} =0$ for $s_i = 1$. The resulting action profile $a^*$ is a pure Nash equilibrium of the game. 
%
%
\end{lemma}
\begin{proof} 
We will use the following equivalent definition in \eqref{NE_best_response} of Nash equilibrium for the stage game in \eqref{utility}, 
\begin{equation}
a_i^{*} = \bbone\left({c_0} >  c_1 (1-s_i) \sum_{j\in \ccalN_i} a_j^{*} s_j + { c_2 s_i \sum_{j\in\ccalN_i} a_j^{*} (1-s_j)}\right). \label{fixed_point_NE}
\end{equation}

Note that at the end of the elimination process in Definition \ref{iterated_elimination_process}, the action space of an individual $i$ is either a singleton or is equal to $A_i^* = [0,1]$. A strictly dominated action cannot satisfy the above equation, hence the equilibrium action of an individual with a singleton non-dominated action space is given by $a_i^* = A_i^*$. 

Now consider $i \in \ccalN$ such that $A_i^* = [0,1]$. Suppose $a_i^*=1$ for a susceptible individual ($s_i = 0$) is not a Nash equilibrium action. Then individual $i$ can deviate and by the above equation it must be that $a_i^*=0$ is a Nash equilibrium action because
\begin{equation}
{c_0} <  c_1 (1-s_i) \sum_{j\in \ccalN_i} a_j^{*} s_j.
\end{equation}
Note that by our assumption $a_j^{*} =0$ for all the infected individuals $s_j =0$ at which $0$ is not a strictly dominated action, i.e., when $A_j = [0,1]$. Hence in the right hand side of the above inequality only individuals that do not have $0$ in their not dominated action set will matter. Furthermore, if $0$ is a strictly dominated action, socialize action $1$ is the only remaining not strictly dominated action. Hence, we can write the above inequality as follows,
\begin{equation}
{c_0} <  c_1 (1-s_i) \sum_{j\in \ccalN_i: \{0\} \notin A_j^*} s_j.
\end{equation}
Now note that if the above inequality is true then $1$ should be strictly dominated by action $0$ for individual $i$. Hence, it is a contradiction to the fact that the action space is not a singleton, $A_i^* \in [0,1]$. 

A similar contradiction argument can be made for the infected individuals $s_i = 1$ and the equilibrium action $a_i^* = 0$. Therefore, the action profile described in the statement must be a pure Nash equilibrium of the game. 
\end{proof}

Lemma \ref{iterated_elimination_lemma} shows that the process in Definition \ref{iterated_elimination_process} eliminates all dominated actions in finite time. Furthermore, if all individuals are included in the process, i.e., if $\ccalN = \bigcup_{k=1}^n n(k)$ then we end up with a singleton action profile that is not strictly dominated. This means the game has an unique pure Nash equilibrium by definition of strictly dominated action. If at the end of the process in Definition \ref{iterated_elimination_process}, if all individuals are not included in the process, i.e., $\ccalN \setminus  \bigcup_{k=1}^n n(k) \neq \emptyset$ then the set of not strictly dominated actions is not a singleton. Lemma \ref{NE_action_profile} proposes a pure strategy profile that is a Nash equilibrium of the game in \eqref{utility} for the case that actions spaces of individuals that survive strict elimination process are not all singleton. 

Lemmas \ref{iterated_elimination_lemma} and \ref{NE_action_profile} considered the stage game with payoffs \eqref{utility} given state $s\in \{0,1\}^n$. An MMPE strategy profile $\sigma$ is a mapping from any state to the action space. We can obtain a pure MMPE strategy profile using Lemma \ref{NE_action_profile} for all possible states $s\in \{0,1\}^n$. That is, we use the elimination process in Definition \ref{iterated_elimination_process} and the action assignment given in Lemma \ref{NE_action_profile} for all the states to construct an MMPE strategy profile. In our simulations, in this paper, we construct the MMPE equilibrium strategy profiles following this process -- see Fig. \ref{fig:algorithm_example} for an example. 

\section{Price of Anarchy of the stage game} \label{poa_bound_proof}
In this section, we consider the sub-optimality of the decisions of individuals that play according to an MMPE strategy profile. By the definition of MMPE strategy, individuals consider current payoffs and play according Nash equilibrium strategy of that stage game. In the following we show the worst stage game Nash equilibrium strategy can be $n$ fold worse than the optimal action profile given a network and state.


Define the stage game $\Gamma(s, \ccalG, c)$ with payoffs in \eqref{utility} parametrized by the disease state $s$, contact network $\ccalG$ and utility constants $c := \{c_0, c_1, c_2\}$.  The welfare value of the action profile $a$ in this game is the sum of utilities of the individuals,
\begin{equation}\label{welfare_function}
W(a,s(t)) = \sum_{i =1}^n u_i(a_i, a_{\ccalN_i}, s(t)) = \sum_{i=1}^{n} {c_0}  a_i  -  (c_1 + c_2) \sum_{(i,j) \in \ccalE} (1-s_i(t)) s_j(t) a_i a_j .
\end{equation}
Note that the first summation is over all the individuals and the second summation is over all the edges in the network $\ccalG$. 
We define the optimum action profile as the maximizer of the welfare function above, i.e., $a^{opt} = \argmax_{a} W(a,s(t))$. We denote the set of Nash equilibrium action profiles by $A^{*}$ and define the price of anarchy (PoA) as the ratio of the worst possible Nash action profile to the optimum action profile, 
\begin{equation}\label{price_anarchy_definition}
PoA := \frac{\min_{a \in A^{*}} W(a)}{W(a^{opt})}. 
\end{equation}
Obviously, $PoA \leq 1$. In the following we provide a lower bound on the price of anarchy. 
\begin{proposition}\label{price_anarchy_bound_proposition}
For a game $\Gamma(s, \ccalG, c)$, the price of anarchy \eqref{price_anarchy_definition} has the following lower bound
\begin{equation} \label{price_anarchy_bound}
PoA \geq 1 - \frac{\max_{i \in \ccalN} |\ccalN_i| \max(c_1,c_2)}{n c_0} 
\end{equation}
where $|\ccalN_i|$ is the degree connectivity of $i$.
\end{proposition}
\begin{proof}
Consider the derivatives of $u_i$ in \eqref{utility} and $W$ in \eqref{welfare_function} with respect to $a_i$, respectively, 
\begin{align}
\frac{\partial u_i}{\partial a_i} &=  c_0 -  \bigg(c_1 (1-s_i) \sum_{j \in \ccalN_i} s_j a_j + c_2 s_i \sum_{j \in \ccalN_i} (1-s_j)a_j\bigg) \\
\frac{\partial W}{\partial a_i} &=  c_0 -  (c_1 + c_2) \bigg((1-s_i) \sum_{j \in \ccalN_i} s_j a_j + s_i \sum_{j \in \ccalN_i} (1-s_j)a_j \bigg). 
\end{align}
First note that $\frac{\partial u_i}{\partial a_i}  \geq \frac{\partial W}{\partial a_i}$ for any $s$ and $a_{-i}$ because 
\begin{equation} \label{partial_W_i}
\frac{\partial W}{\partial a_i} = \frac{\partial u_i}{\partial a_i} - \bigg(c_2 (1-s_i) \sum_{j \in \ccalN_i} s_j a_j + c_1 s_i \sum_{j \in \ccalN_i} (1-s_j)a_j\bigg).
\end{equation}
Therefore, given actions of others $a_{-i}$ and state $s$, it could be that $\frac{\partial u_i}{\partial a_i}>0$ while $\frac{\partial W}{\partial a_i}<0$. Consider the optimal and equilibrium action profiles where $a_{j}^{opt} = a_{j}^{*}$ for $j \in \ccalN\setminus i$, and $a_i^{opt} = 0$ but $a_i^{*} = 1$. That is, we have  $\frac{\partial u_j}{\partial a_j}<0$ and $\frac{\partial W}{\partial a_j}<0$ for all $j \neq i$ but $\frac{\partial u_i}{\partial a_i}>0$ while $\frac{\partial W}{\partial a_i}<0$. Then, the difference between the equilibrium profile welfare $W(a^{*},s)$ and optimal action profile welfare $W(a^{opt},s)$ is equal to \eqref{partial_W_i}, that is,
\begin{equation} \label{difference_welfare}
W(a^{*},s) - W(a^{opt},s) = \frac{\partial W}{\partial a_i} 
\end{equation}
We divide the difference above by $W(a^{opt},s)$ to get 
\begin{equation}
-1 +  \frac{W(a^{*},s)}{W(a^{opt},s)} = \frac{\frac{\partial W}{\partial a_i}}{W(a^{opt},s)}.
\end{equation}
Now assume $\frac{\partial u_i}{\partial a_i} = \epsilon>0$ to get 
\begin{equation}
  \frac{W(a^{*},s)}{W(a^{opt},s)} = 1 + \frac{\epsilon - \bigg(c_2 (1-s_i) \sum_{j \in \ccalN_i} s_j a_j + c_1 s_i \sum_{j \in \ccalN_i} (1-s_j)a_j\bigg)}{W(a^{opt},s)}.
\end{equation}
A trivial lower bound on the term for the numerator above is $-\max(c_1, c_2) |\ccalN_i|$ for any state $s$. Furthermore a trivial upper bound for $W(a^{opt},s)$ is $n c_0$. Using these bounds we obtain a bound for the above equality,
\begin{equation}
\frac{W(a^{*},s)}{W(a^{opt},s)} \geq 1 - \frac{\max(c_1, c_2) |\ccalN_i|}{n c_0}
\end{equation}
Now note that the choice of the individual $i$ is arbitrary. When this individual is the individual with the largest number of neighbors we get a lower bound on the right hand side of the above inequality which is the worst case lower bound on the price of anarchy in \eqref{price_anarchy_bound}.  
\end{proof}

Note that this upper bound can be arbitrarily bad, i.e., in the order of $1/n$. In the discussion following Figure 2(d) in the main text we show that this bound is tight.  We repeat this example below.

\begin{figure}[t]
\centering
\begin{tabular}{cc}
	\includegraphics{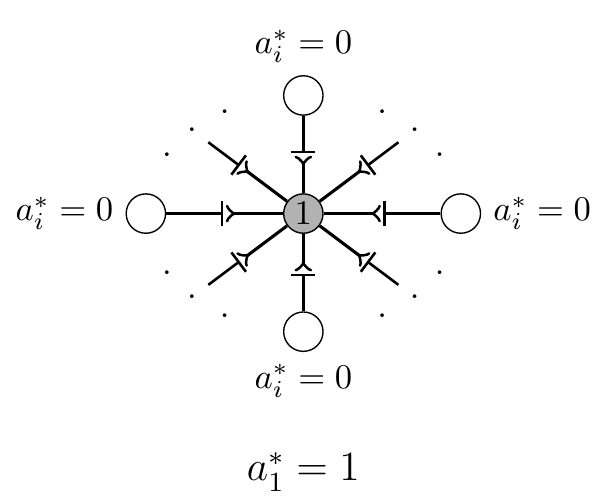}&
	\includegraphics{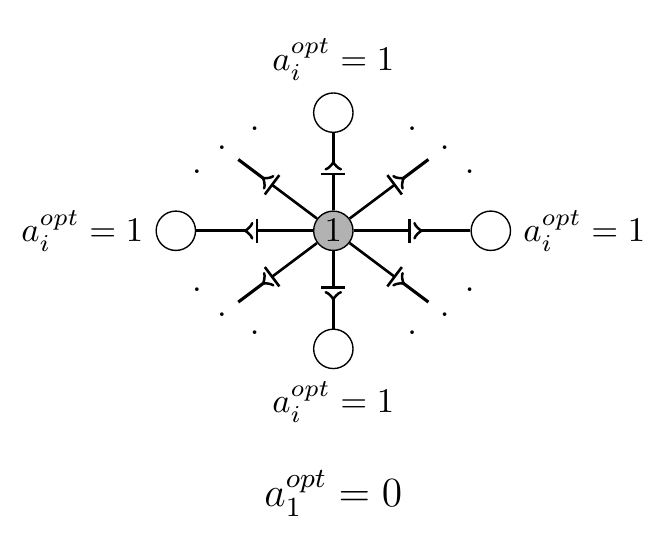}
\end{tabular}
	\caption{\small Price of anarchy in the star network (Example \ref{star_network_example}). We consider the strong empathy ($c_0<(n-1) c_2$) and strong averseness  ($c_0<c_1$) case in Figure 2(d) of the main manuscript. Left and right figures show the two stage game pure Nash equilibria. The Nash equilibrium action profile shown by the right figure is optimal with welfare equal to $(n-1)c_0$. The Nash equilibrium action profile shown by the left figure attains a welfare equal to $c_0$. The PoA is equal to $1/n-1$. }
	\label{fig:poa_example}
\end{figure}

\begin{example}\label{star_network_example}
Consider a star network with $n$ individuals. The center individual with $n-1$ neighbors is the only infected individual. Payoff constants are such that $c_0< c_1$ and $c_0 < (n-1) c_2$. There are two stage game Nash equilibria: 1) $a_i^{opt} = 1$ if $s_i = 0$ and $a_i^{opt} = 0$ if $s_i = 1$, and 2) $a_i^{*} = 0$ if $s_i = 0$ and $a_i^{*} = 1$ if $s_i = 1$. The first Nash equilibrium $a^{opt}$ is also the optimal action profile yielding a welfare of $(n-1)c_0$. The second Nash equilibrium $a^{*}$ obtains a welfare of $c_0$ resulting in a $PoA$ of $\frac{1}{n-1}$. 
\end{example}

Another example on a star network follows. 

\begin{example}\label{star_network_example_2}
Consider the star network shown in Figure \ref{fig:poa_example}. The center individual 1 has $n-1$ neighbors and is the only infected individual. Constants are such that $c_0 = c_1 + \rho_1$ and $c_0 = (n-1) c_2+\rho_2$ for arbitrarily small positive constants $\rho_1>0$ and $\rho_2>0$, so that $c_0 < (n-1) (c_1 + c_2)$. The unique NE  action profile is that all individuals are social, $a_i^{*} = 1$ for all $i \in \ccalN$. The welfare for the NE is $W(a^{*}) = n c_0 - (n-1)(c_1+c_2) = (n-1) \rho_1 + \rho_2$. The optimal action profile is that only the susceptible nodes are social $a_i^{opt} = 1$ for $i \neq 1$ and $a_1^{opt} = 0$. The welfare for the $a^{opt}$ is $W(a^{opt}) = (n-1) c_0$. Then $PoA \approx \rho_1 / c_0$ for large $n$. 
\end{example}
Here, we see that $PoA$ is determined by the closeness of $c_1$ to $c_0$. Note that in this example, equilibrium is unique and, unlike Example \ref{star_network_example}, the optimal action profile is not a Nash equilibrium.  

Another notion that enables us to gauge the optimality of equilibria is the Price of Stability (PoS). PoS is the ratio of the best possible Nash action profile to the optimum action profile, 
\begin{equation}\label{price_stability_definition}
PoS := \frac{\max_{a \in A^{*}} W(a)}{W(a^{opt})}
\end{equation}
Note that $PoS\leq 1$ by definition of $a^{opt}$. 
We remark that in Example \ref{star_network_example_2}, the equilibrium is unique, meaning $PoS = PoA$. Therefore, not only the $PoA$ but also the $PoS$ can be arbitrarily bad.

\section{$R_0$ bound} \label{r_0_bound_proof}

We formally define the reproductive ratio $R_0$ in the following. 
\begin{definition}\label{R_0_definition}
Let the initial state of the population be given by $s(0)$ where $s_i(0) = 1$, otherwise $s_j(0) = 0$ for all $j\neq i$ for some randomly selected individual $i$. Then $R_0$ is the expected number of individuals that contract the disease from the randomly selected individual $i$ until $i$ heals,
\begin{equation}\label{eq:R_0_definition}
R_0 := E\left[ E\left[\sum_{t=1}^\infty \sum_{j=1}^n \bbone(s_j(t+1) - s_j(t) = 1, i \to j) \bbone\left(s_i(l) = 1 \textrm{ for } l <t\right) \given s(0)  \right] \right]
\end{equation}
where $\bbone(s_j(t+1) - s_j(t) = 1, i \to j)$ is the indicator function that is one if individual $j$ transitions to an infected state at time $t$ and $i$ is the one infecting $j$, and $\bbone\left(s_i(l) = 1 \textrm{ for } l <t\right)$ is the indicator function that is one if individual $i$ has not healed yet. The outside expectation is with respect to the uniform distribution that selects the initial infected individual, and the inside expectation is with respect to the transition probabilities of the Markov chain. 
\end{definition}
%
In the following, we derive bounds for $R_0$ given that individuals act according to an MMPE strategy profile and we select the initial infected individual randomly from the network. 
\begin{theorem} \label{theorem_r_0_bound_generic}
Consider a network with degree distribution $P(k)$. Assume the infected individual is chosen  from the population uniformly at random. Assume $c_0>c_1$. Then $R_0$ defined in \eqref{eq:R_0_definition} has the following upperbound,
\begin{equation} \label{r_0_bound}
R_0 \leq \frac{\beta}{\delta}\sum_{k=1}^{K} k P(k)
\end{equation}
where $K := \min(\lfloor c_0/c_2 \rfloor, n)$.
\end{theorem}
\begin{proof}
We start by moving the second expectation inside the sum in the definition of $R_0$ \eqref{eq:R_0_definition} to get the following, 
\begin{equation} \label{eq:R_0_expectation_inside}
R_0 =  E\left[ \sum_{t=1}^\infty  \sum_{j=1}^n E\left[ \bbone(s_j(t+1) - s_j(t) = 1, i \to j)  \bbone\left(s_i(l) = 1 \textrm{ for } l <t\right) \given s(0)  \right] \right].
\end{equation}
Now consider the conditional expectation inside the summation which we can equivalently represent as the following conditional probability
\begin{align} \label{eq:expectation_to_probability}
E[ \bbone(s_j(t+1) - s_j(t) = 1&, i \to j)  \bbone\left(s_i(l) = 1 \textrm{ for } l <t\right) \given s(0)  ] \nonumber \\
&= P\left(s_j(t+1) -s_j(t) = 1, i \to j, s_i(l) = 1 \textrm{ for } l <t \given s(0)\right).
\end{align}
Note that the above conditional probability is the probability that $j$ is infected by $i$ at time $t$ and $i$ remained infected until time $t$ given $i$ is infected at $t=0$. Using the chain rule and law of total probability, we can write the above conditional probability as follows
\begin{align}
P&(s_j(t+1) -s_j(t) = 1, i \to j, s_i(l) = 1 \textrm{ for } l <t \given s(0)) = \nonumber \\
&\bigg(P\left(s_j(t+1) -s_j(t) = 1, i \to j\given s_i(t) = 1,s_j(t) = 0,s(0)\right)  P(s_i(t) = 1,s_j(t) = 0 \given s_i(l) = 1 \textrm{ for } l <t,s(0) )\nonumber \\
&+P\left(s_j(t+1) -s_j(t) = 1, i \to j\given s_i(t) = 0,s_j(t) = 0,s(0)\right)  P( s_i(t) = 0,s_j(t) = 0 \given s_i(l) = 1 \textrm{ for } l <t,s(0) )\nonumber \\
&+P\left(s_j(t+1) -s_j(t) = 1, i \to j\given s_i(t) = 1,s_j(t) = 1,s(0)\right)  P(s_i(t) = 1,s_j(t) = 1 \given s_i(l) = 1 \textrm{ for } l <t,s(0) )\nonumber \\
&+P\left(s_j(t+1) -s_j(t) = 1, i \to j\given s_i(t) = 0,s_j(t) = 1,s(0)\right)  P( s_i(t) = 0 ,s_j(t) = 1 \given s_i(l) = 1 \textrm{ for } l <t,s(0) )\bigg)\nonumber \\
&P\left(s_i(l) = 1 \textrm{ for } l <t \given s(0)\right)  
\end{align}
Note that the first four lines equal to the probability that $i$ infects $j$ at time $t$ given that $i$ remained infected until $t-1$ by law of total probability. The last line is the probability that $i$ remained infected given that $i$ started infected since $s_i(0) = 1$. Observe that the probability that $i$ infects $j$ is zero if individual $i$ is susceptible at time $t$ or individual $j$ is infected, i.e., $P\left(s_j(t+1) -s_j(t) = 1, i \to j\given s_i(t) = 0\right) = 0$ or $P\left(s_j(t+1) -s_j(t) = 1, i \to j\given s_j(t) = 1\right) = 0$. Hence only the first line of the four line expression is nonzero which simplifies the identity above as follows
\begin{align} \label{conditional_probability_j_infects_i}
P(&s_j(t+1) -s_j(t) = 1, i \to j, s_i(l) = 1 \textrm{ for } l <t \given s(0)) = \nonumber \\
&P\left(s_j(t+1) -s_j(t) = 1, i \to j\given s_i(t) = 1,s_j(t) = 0,s(0)\right)  P(s_i(t) = 1, s_j(t) = 0 \given s_i(l) = 1 \textrm{ for } l <t ,s(0))\nonumber \\
&P\left(s_i(l) = 1 \textrm{ for } l <t \given s(0)\right)  
\end{align}
The probability of healing at each step is independent, hence the last conditional probability is equal to $(1-\delta)^{t-1}$. The conditional probability that $i$ remains infected and $j$ is infected at time $t$ given that $i$ is infected until time $t$ is less than $1-\delta$ by the argument below,
\begin{align}
P(&s_i(t) = 1,  s_j(t) = 0 \given s_i(l) = 1 \textrm{ for } l <t ,s(0))  \nonumber \\
&= P(s_j(t) = 0 \given  s_i(t) = 1,  s_i(l) = 1 \textrm{ for } l <t,s(0) )  P(  s_i(t) = 1 \given s_i(l) = 1 \textrm{ for } l <t,s(0) ) \\
& \leq P(  s_i(t) = 1 \given s_i(l) = 1 \textrm{ for } l <t,s(0)) \\
& = 1-\delta. \;\; 
\end{align}
The first equality above follows by chain rule. The inequality follows by the fact that the probability is less than one. The second equality is true by the transition probability of individual $i$ from an infected state. Substituting these identities inside the equation \eqref{conditional_probability_j_infects_i} we obtain the following 
\begin{align} \label{eq:contraction_probability}
P(s_j(t+1)& -s_j(t) = 1, i \to j, s_i(l) = 1 \textrm{ for } l <t \given s(0)) \leq \nonumber \\
&P\left(s_j(t+1) -s_j(t) = 1, i \to j\given s_i(t) = 1,s_j(t) = 0,s(0)\right)   (1-\delta)^{t}.
\end{align}

Now consider the conditional probability on the right hand side of \eqref{eq:contraction_probability}, the probability that $j$ is infected at time $t+1$ by $i$ given that $j$ is susceptible and $i$ is infected at time $t$. We have the following upper bound,
\begin{align} 
P(s_j(t+1) &-s_j(t) = 1, i \to j \given s_i(t) = 1,s_j(t) = 0,s(0)) \nonumber \\
&=  P\left(s_j(t+1) = 1, i \to j\given s_i(t) = 1,s_j(t) = 0,s(0)\right) \nonumber \\
& = P\left(s_j(t+1) = 1, i \to j \given a_i^{*}(t) = 1, a_j^{*}(t)=1, s_i(t) = 1,s_j(t) = 0,s(0)\right)\nonumber \\
& \qquad P\left(a_i^{*}(t) = 1, a_j^{*}(t) =1 \given s_i(t) = 1,s_j(t) = 0,s(0)\right)
 \nonumber \\
 & \leq P\left(s_j(t+1) = 1, i \to j \given a_i^{*}(t) = 1, a_j^{*}(t)=1, s_i(t) = 1,s_j(t) = 0,s(0)\right)\nonumber \\
 & \qquad P\left(a_i^{*}(t) = 1 \given s_i(t) = 1,s_j(t) = 0,s(0)\right) \nonumber \\
 & = \beta \bbone(j\in \ccalN_i) P\left(a_i^{*}(t) = 1 \given s_i(t) = 1,s_j(t) = 0,s(0)\right)\label{eq:upper_bound_infection_j} 
\end{align}
The first equality is by the fact that $s_j(t+1) = 1$ if $s_j(t)=0$ and $s_j(t+1)-s_j(t) = 1$. The second equality above is by the law of total probability and by the fact that if $i$ or $j$ takes an action to self-quarantine, i.e., $a_j(t)=0$ or $a_i(t)=0$, then $i$ cannot infect $j$. The inequality follows by the fact that $P\left(a_i(t) = 1, a_j(t) =1 \given s_i(t) = 1,s_j(t) = 0\right)  \leq P\left(a_i(t) = 1 \given s_i(t) = 1,s_j(t) = 0\right)$. The last equality follows because if both agents socialize at normal levels then the infection probability is $\beta$ when agent $j$ and $i$ are connected, $j \in \ccalN_i$. 

Now we consider $a_i^{*}(t)$. We know that the MMPE action of individual $i$ is a best response to best response actions of neighbors from \eqref{NE_best_response} where best response function is given by \eqref{best_response_specific}. Specifically from the perspective of an infected individual $i$, its best response action is given by the following
\begin{equation}
a_i^{*}(t) =  \bbone\left({c_0} > { c_2  \sum_{j\in\ccalN_i} a_j^{*}(t) (1-s_j(t))}\right) \label{best_response_specific_infected}
\end{equation}
Consider $t=1$. If $c_0 >  c_1$, then $a_j^{*}(1) = 1$ for all $j \in \ccalN_i$. Hence, if $c_0 >  c_2 |\ccalN_i|$ then $a_i^{*}(1) = 1$, i.e., $a_i^{*}(1) = \bbone(c_0 >  c_2 |\ccalN_i|)$ . Also note that if $a_i^{*}(1) = 1$ then $a_i^{*}(t) = 1$ for all $t >1$. Moreover, if $a_i^{*}(1) =0$ then $i$ never infects another node if it is the only initially infected individual. Hence, we have $a_i^{*}(t) = a_i^{*}(1)$ for all $t$. That is, the action of agent $i$ at time $t$ is determined by its initial action and is independent of the state at time $t$. Therefore, we can write
\begin{align}
P\big(a_i^{*}(t) = 1 \given s_i(t) = 1,s_j(t) = 0,s(0)\big)& = P\left(a_i^{*}(1) = 1 \given s(0)\right)\nonumber \\
& = \bbone(c_0 >  c_2 |\ccalN_i|).
\end{align}
Substituting the above identity in \eqref{eq:upper_bound_infection_j} and using \eqref{eq:upper_bound_infection_j} in \eqref{eq:contraction_probability}, which then we substitute to \eqref{eq:expectation_to_probability}, we get the following upper bound for $R_0$ in \eqref{eq:R_0_expectation_inside}, 
%
%
%
%
\begin{align}
R_0 &\leq E\left[ \sum_{t=1}^\infty  \sum_{j=1}^n \beta \bbone(c_0 >  c_2 |\ccalN_i|)  \bbone(j \in \ccalN_i) (1-\delta)^{t} \right].
\end{align}
Now note the indicator function terms depend on the network of individual $i$ but they do not depend on the identities of its neighbors. In fact, $\sum_{j=1}^n \bbone(j \in \ccalN_i)$ is equal to the number of neighbors of $i$, i.e., $|\ccalN_i|$. In addition, we have $\sum_{t=1}^\infty  (1-\delta)^{t} = 1/\delta$.  Using these identities, we can write the above bound as follows,  
\begin{align} \label{eq:bound_with_sampling_expectation}
R_0 &\leq \frac{\beta}{\delta}E\left[  |\ccalN_i|  \bbone(c_0 >  c_2 |\ccalN_i|)  \right] \end{align}
Individual $i$ is chosen uniformly random and it has $k$ neighbors with probability $P(k)$, therefore, the expectation on the right hand side is equal to the following,
\begin{align}
R_0 &\leq \frac{\beta}{\delta}\sum_{k=1}^n   k  \bbone( k<\frac{c_0}{ c_2 })   P(k)
\end{align}
Bound in \eqref{r_0_bound} follows. 
\end{proof}
%

\section{$R_0$ bound scale-free} \label{r_0_bound_proof}
\begin{corollary} \label{theorem_r_0_bound}
Consider a random scale free network where the degree distribution follows $P(k) \sim k^{-\gamma}$ for $\gamma = 2$. Then $R_0$ defined in \eqref{eq:R_0_definition} has the following upper bound,
\begin{equation} \label{r_0_bound_scale_free}
R_0 \leq \frac{n}{2 n -1}\frac{\beta}{\delta} \log(\min(\lfloor c_0/c_2 \rfloor, n)+1).
\end{equation}
\end{corollary}
\begin{proof}
We directly use the bound in \eqref{r_0_bound} and substitute in $P(k) = L(-2,n) k^{-2}$ where $L(-2,n) = (\sum_{k=1}^n k^{-2})^{-1}$ is the normalization constant for the scale-free distribution
\begin{align} \label{r_0_bound_scale_free_v1}
R_0 \leq  L(-2,n) \frac{\beta}{\delta}\sum_{k=1}^{\min(\lfloor c_0/c_2 \rfloor, n)} k^{-1}
\end{align}
We first upper bound the normalization constant $L(-2,n)$ by the fact that $\sum_{k=1}^n k^{-2} > 2 - \frac{1}{n}$. Hence, $L(-2,n) \leq \frac{n}{2n-1}$. Next we note that the summation behaves logarithmically which yields the desired bound in \eqref{r_0_bound_scale_free}.
\end{proof}

\section{$R_*$ bound} \label{r_x_bound_proof}

We formally define the reproductive ratio $R_*$ in the following. 
\begin{definition}\label{R_x_definition}
Let the initial state of the population be given by $s(0)$ where $s_i(0) = 1$, otherwise $s_j(0) = 0$ for all $j\neq i$ for some randomly selected individual $i$. The probability that we select an initial sick individual with degree $k$ is given by $Q(k) := \frac{k P(k)}{\sum_k k P(k)}$. Then $R_*$ is the expected number of individuals that contract the disease from a randomly selected individual $i$ until $i$ heals,
\begin{equation}\label{eq:R_x_definition}
R_* := E\left[ E\left[\sum_{t=1}^\infty \sum_{j=1}^n \bbone(s_j(t+1) - s_j(t) = 1, i \to j) \bbone\left(s_i(l) = 1 \textrm{ for } l <t\right) \given s(0)  \right] \right]
\end{equation}
where $\bbone(s_j(t+1) - s_j(t) = 1, i \to j)$ is the indicator function that is one if individual $j$ transitions to an infected state at time $t$ and $i$ is the one infecting $j$, and $\bbone\left(s_i(l) = 1 \textrm{ for } l <t\right)$ is the indicator function that is one if individual $i$ has not healed yet. The outside expectation is with respect to the probability distribution $Q(k)$, and the inside expectation is with respect to the transition probabilities of the Markov chain. 
\end{definition}

Below we present a bound for $R_*$ for a generic network. 

\begin{theorem} \label{theorem_r_x_bound_generic}
Consider a network with degree connectivity distribution $P(k)$. Scale $c_0$ with $\beta$ for convenience, i.e., $c_0 := \beta c_0$, and assume $c_0>c_1$. Then $R_*$ defined in Definition \ref{R_x_definition} has the following upper bound,
\begin{equation} \label{r_x_bound}
R_* \leq \frac{\beta}{\delta}\sum_{k=1}^{\min(\lfloor c_0/c_2 \rfloor, n)} \frac{k^2 P(k)}{\sum_{k=1}^n k P(k)}
\end{equation}
\end{theorem}
\begin{proof}
First note that the only difference between the $R_0$ definition and $R_*$ is the difference in probability distributions of selecting the initial infected individual. Hence, the bound in \eqref{eq:bound_with_sampling_expectation} applies to $R_*$ which we repeat here for convenience, 
\begin{align} \label{eq:bound_with_sampling_expectation_v1}
R_* &\leq \frac{\beta}{\delta}E\left[  |\ccalN_i|  \bbone(c_0 >  c_2 |\ccalN_i|)  \right] \end{align}
where now the expectation is with respect to the distribution $Q(k)$. Therefore, the expectation on the right hand side is equal to the following,
\begin{align}
R_* &\leq \frac{\beta}{\delta}\sum_{k=1}^n   k  \bbone( k<\frac{c_0}{ c_2 })   Q(k)
\end{align}
Bound in \eqref{r_x_bound} follows by using the definition of $Q(k)=\frac{k P(k)}{\sum_k k P(k)}$.

\end{proof}

\section{$R_*$ bound scale-free} \label{r_x_bound_proof}

\begin{corollary} \label{theorem_r_x_bound_scalefree}
Consider a random scale free network where the degree distribution follows $P(k) \sim k^{-\gamma}$ for $\gamma = 2$. Then $R_*$ given by Definition \ref{R_x_definition} has the following upper bound,
\begin{equation} \label{r_x_bound_scalefree}
R_*\leq \frac{\beta}{\delta} \frac{\min(\lfloor c_0/c_2 \rfloor, n)}{\log(n)}.
\end{equation}
\end{corollary}
\begin{proof}
We directly use the bound in \eqref{r_x_bound} and substitute in $P(k) = L(-2,n) k^{-2}$ where $L(-2,n) = (\sum_{k=1}^n k^{-2})^{-1}$ is the normalization constant for the scale-free distribution
\begin{align} \label{r_x_bound_scale_free_intermediate}
R_* \leq  \frac{\beta}{\delta}\sum_{k=1}^{\min(\lfloor c_0/c_2 \rfloor, n)} \frac{ L(-2,n)}{L(-2,n) \sum_{l=1}^n l^{-1}} = \frac{\beta}{\delta}\sum_{k=1}^{\min(\lfloor c_0/c_2 \rfloor, n)} \frac{ 1}{\sum_{l=1}^n l^{-1}}.
\end{align}
The bound in \eqref{r_x_bound_scalefree} follows by noting that $\sum_{l=1}^n l^{-1} \approx \log(n)$. 
\end{proof}

\begin{figure*}[t]
\centering
\begin{tabular}{ccc}
$\beta = 0.1$ & $\beta = 0.2$ & $\beta = 0.3$ \\
\includegraphics[width=0.32\linewidth]{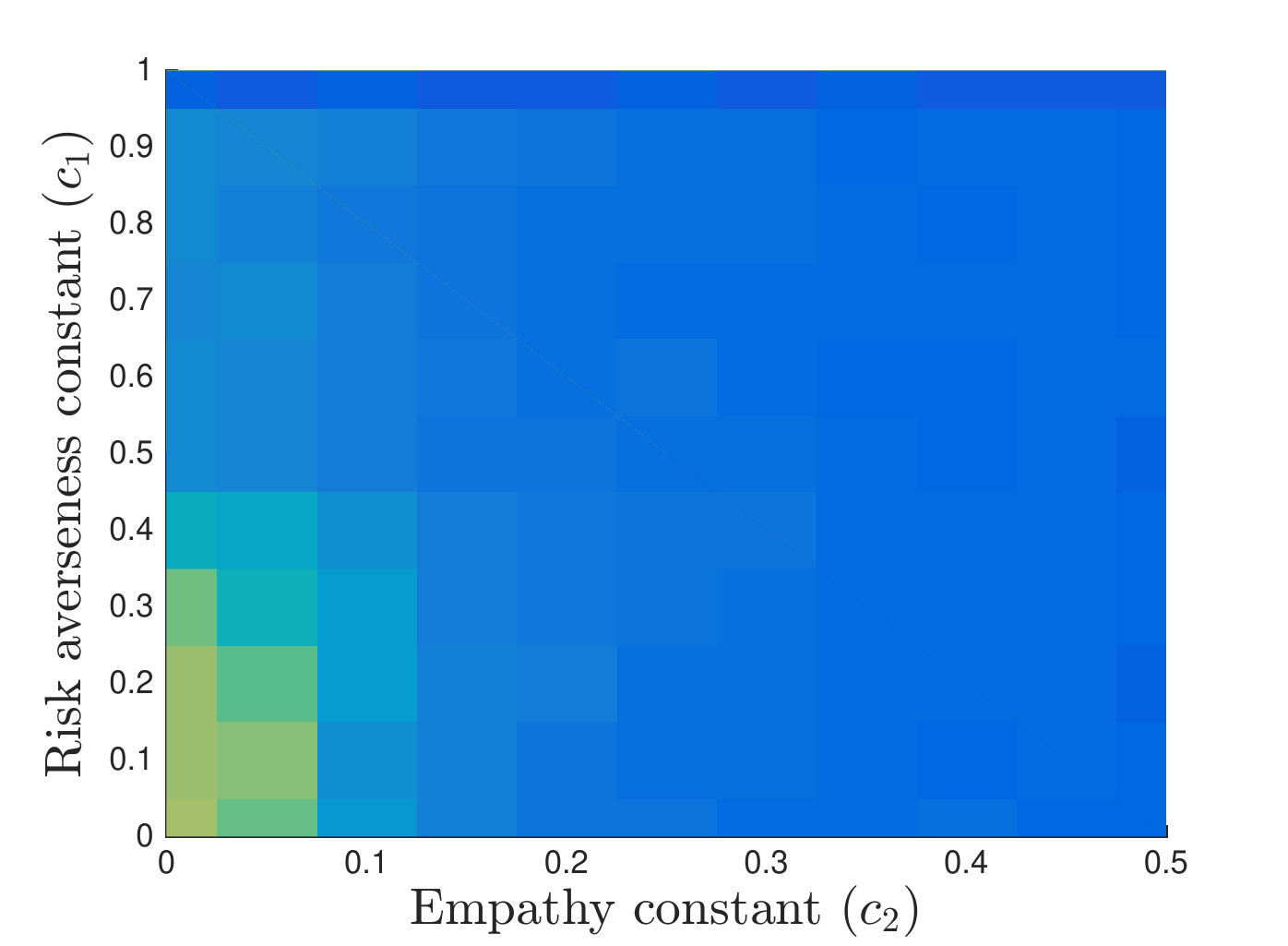}&
\includegraphics[width=0.32\linewidth]{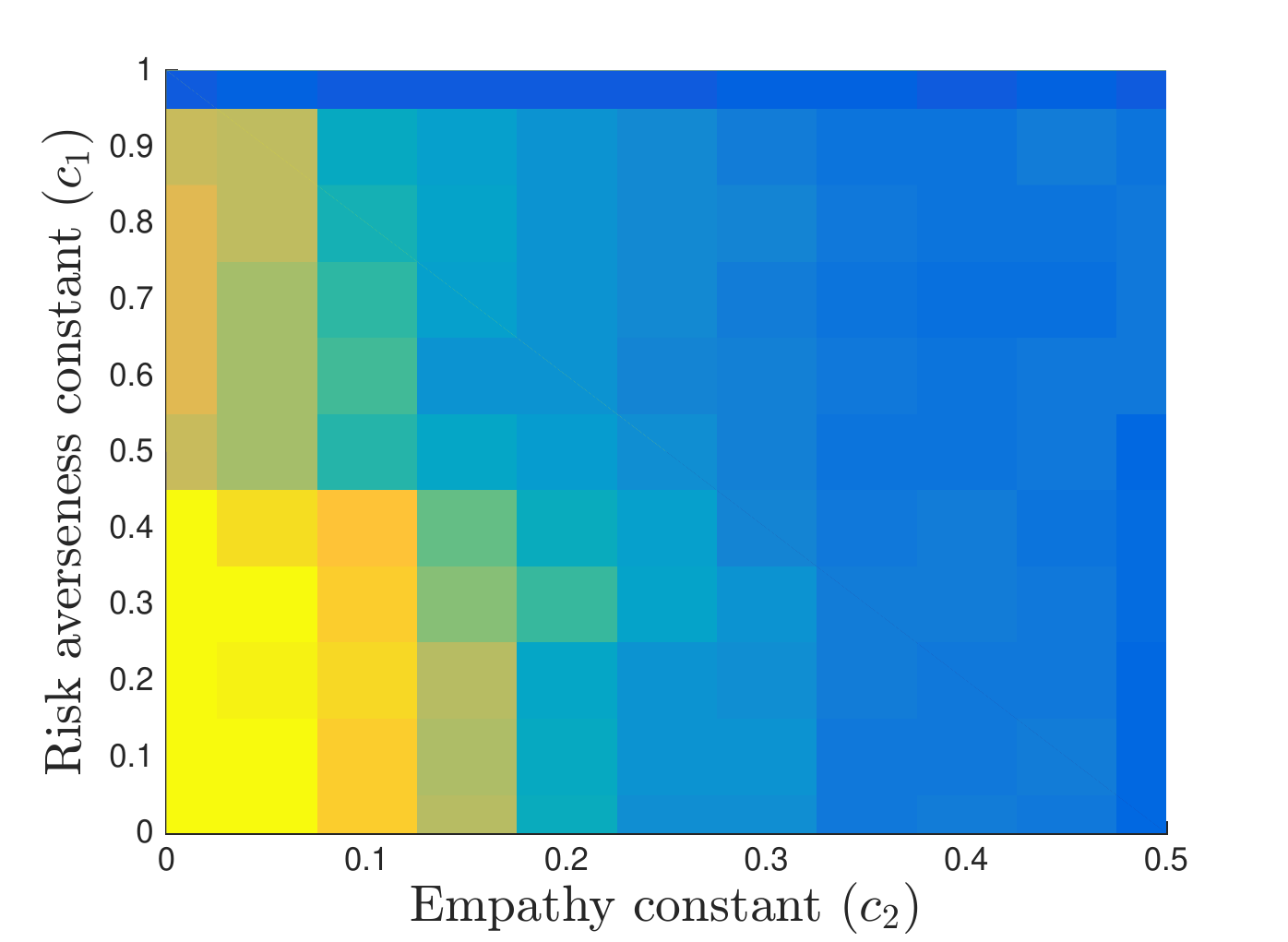}&
\includegraphics[width=0.36\linewidth]{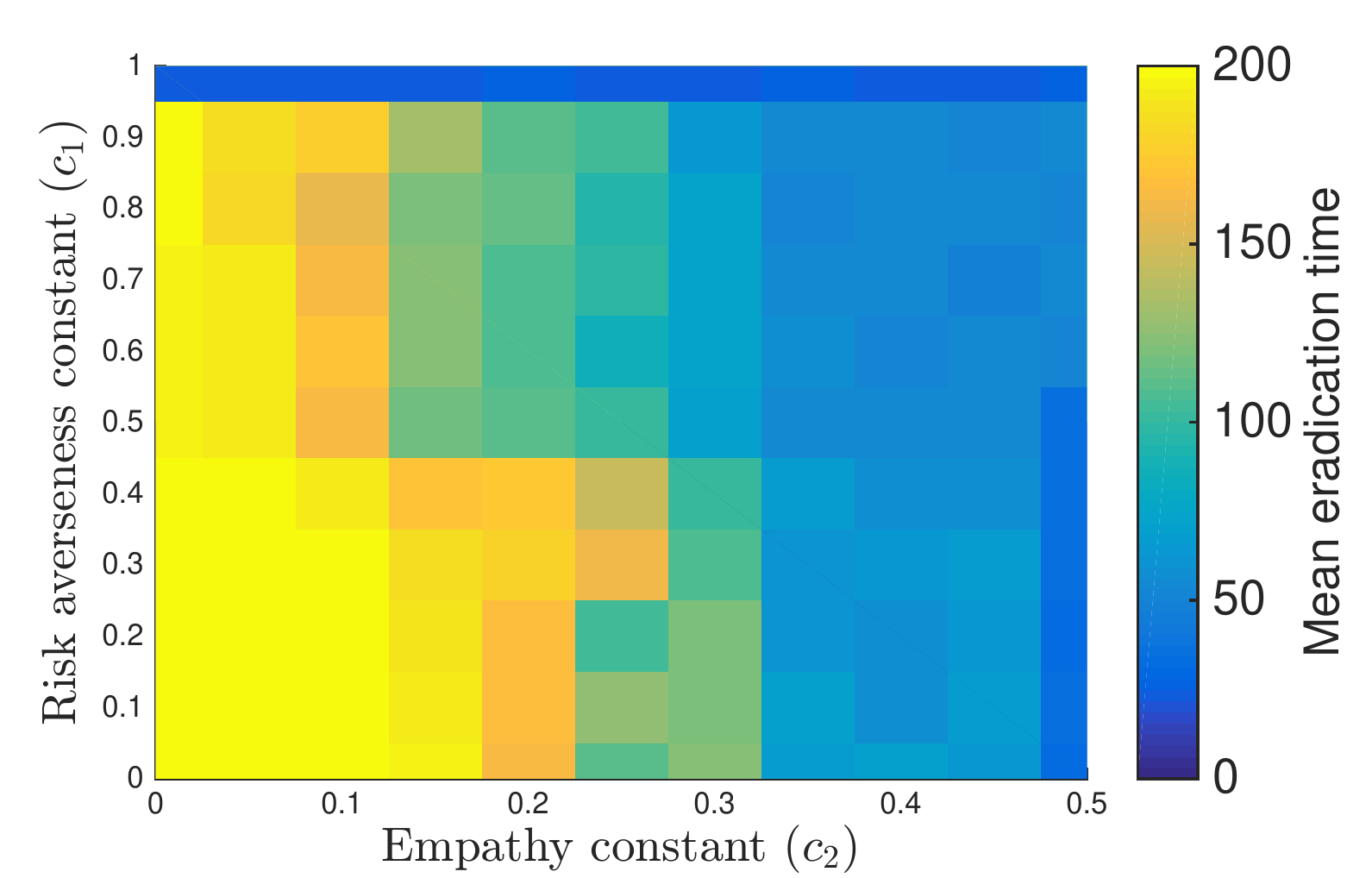}
\end{tabular}
\caption{Effect of risk averseness $c_1$ and empathy $c_2$ constants on mean eradication time. We fix the healing rate to $\delta = 0.2$ and the population size to $n=100$. The infection rate $\beta$ values equal to 0.1,0.2 and 0.3 for figures left, middle and right, respectively. We let $c_0 = \beta$ for each figure. The axes in the figures correspond to the constant values of $c_1$ and $c_2$. For a given value of $c_1$ and $c_2$, we generate 50 scale-free networks using the preferential attachment algorithm and run the stochastic disease network game for $200$ steps for each network. Each run starts with all individuals infected. The grid color represents the average eradication time among runs where we let eradication time be equal to $200$ if the disease is not eradicated. The eradication time decreases as $c_1$ or $c_2$ increases in the region where disease is eradicated fast, i.e., $R_* > 1$. 
}
\label{fig:eradication_time_behavior_beta}
\end{figure*}

\section{Mean eradication time} \label{mean_eradication_time_figures}

See Figure \ref{fig:eradication_time_behavior_beta} for the effect of parameters $c_1$, $c_2$, $\beta$ on eradication time. 

\section{Average infectivity levels}\label{average_infectivity_level_figures}

\begin{figure*}[t]
\centering
\begin{tabular}{ccc}
$\beta = 0.1$ & $\beta = 0.2$ & $\beta = 0.3$ \\
\includegraphics[width=0.32\linewidth]{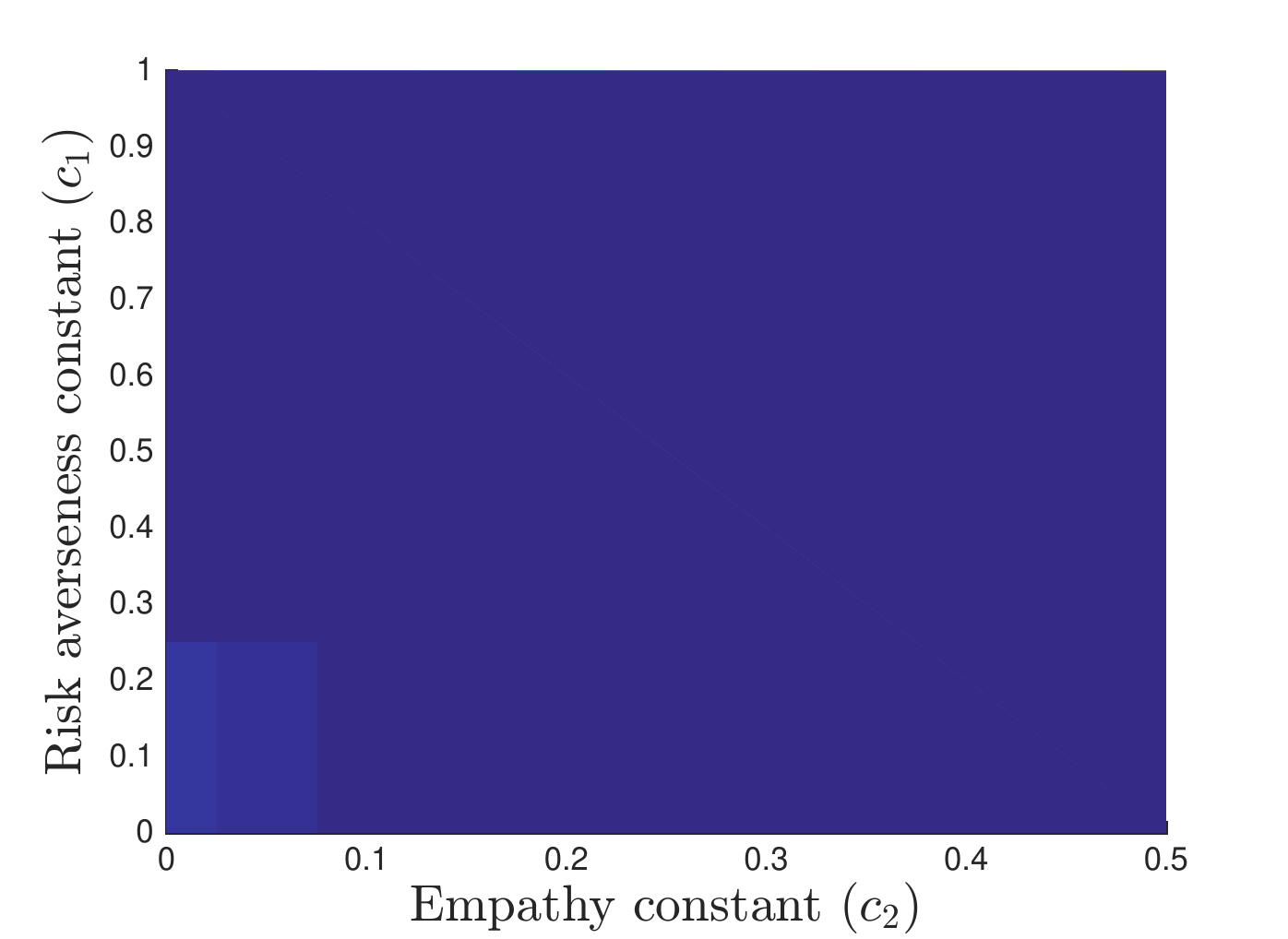}&
\includegraphics[width=0.32\linewidth]{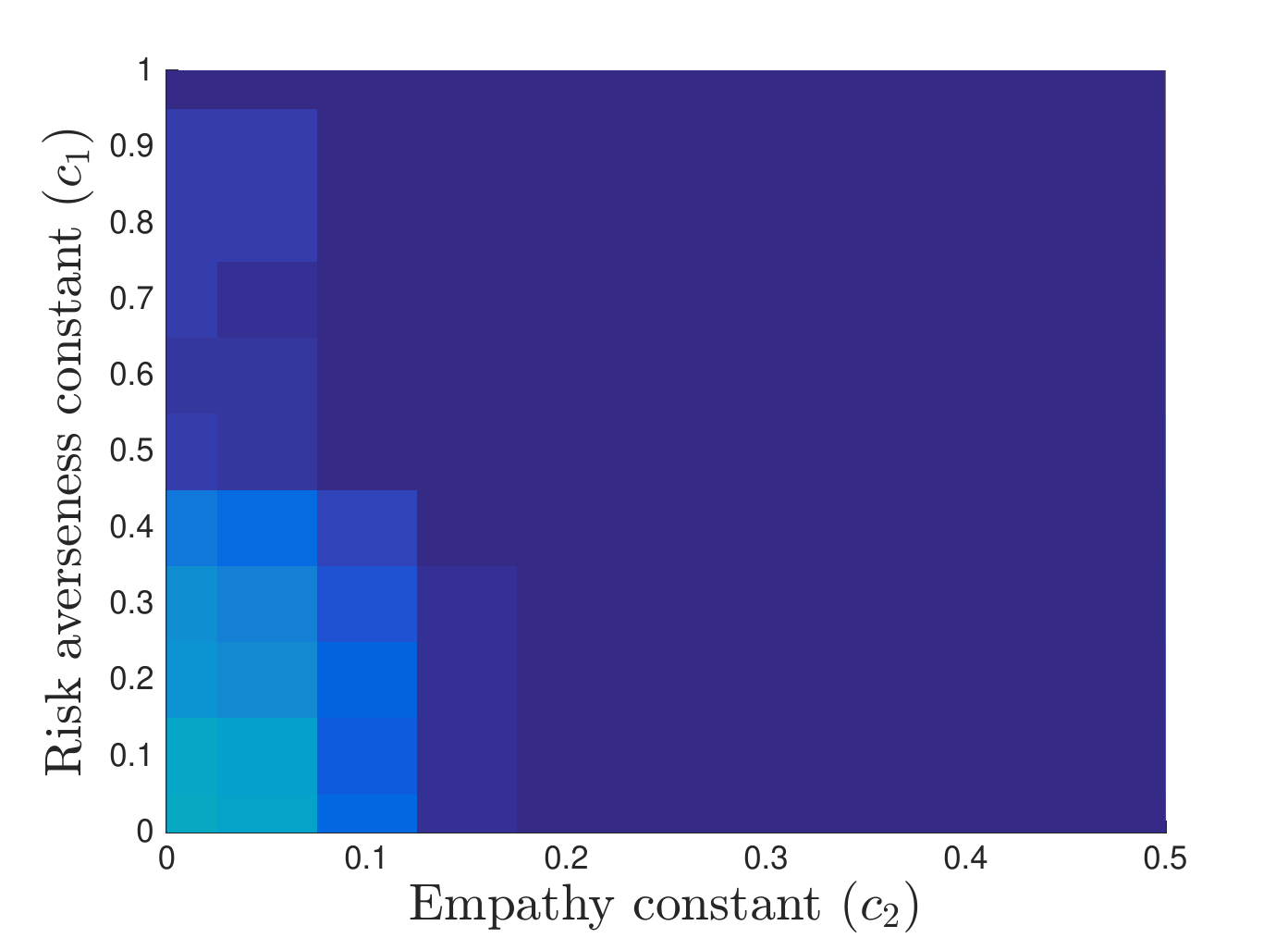}&
\includegraphics[width=0.36\linewidth]{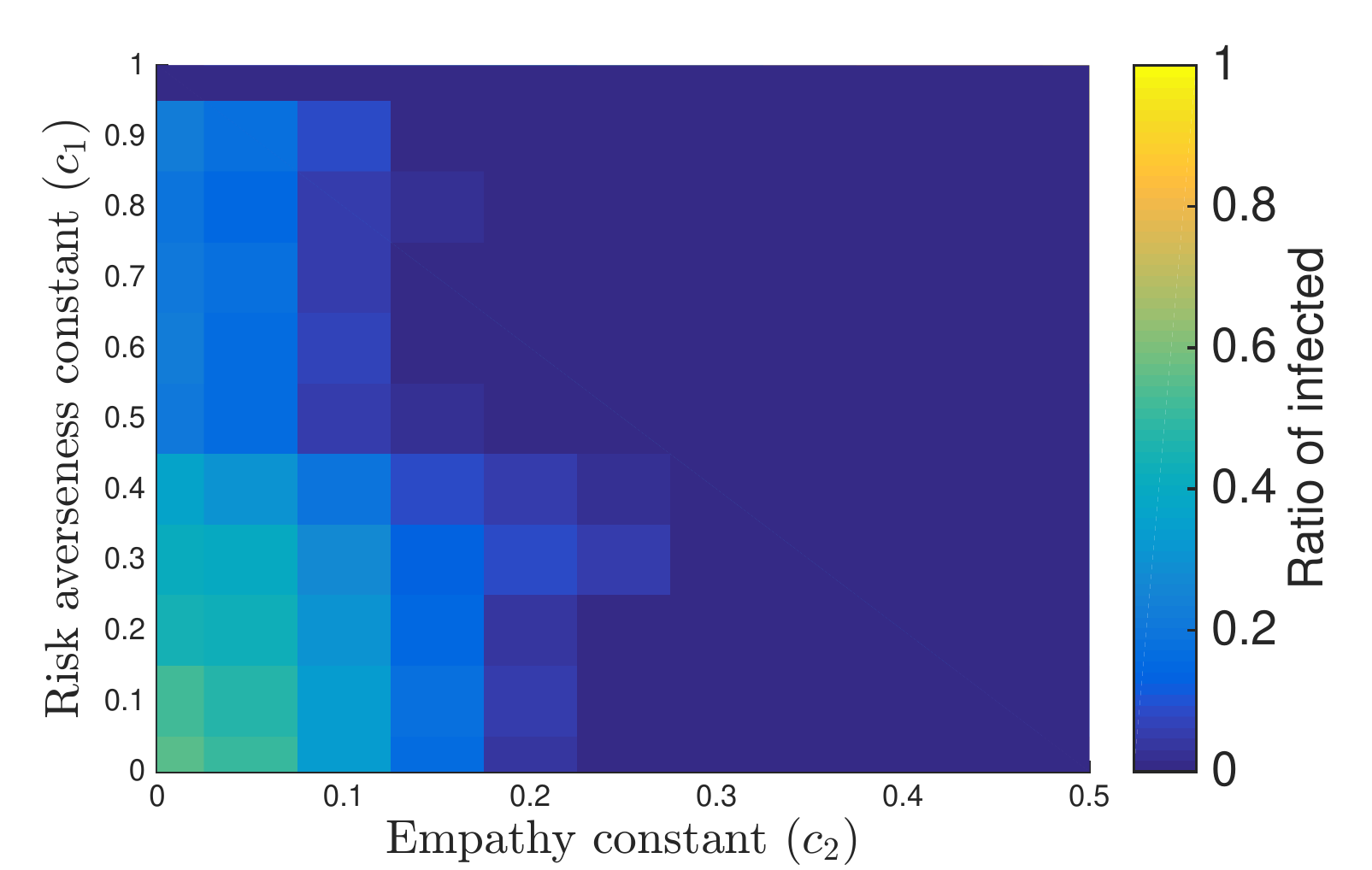}
\end{tabular}
\caption{Effect of risk averseness $c_1$ and empathy $c_2$ constants on average infectivity level. We fix the healing rate to $\delta = 0.2$ and the population size to $n=100$. The infection rate $\beta$ values equal to 0.1,0.2 and 0.3 for figures left, middle and right, respectively. We let $c_0 = \beta$ for each figure. The axes in the figures correspond to the constant values of $c_1$ and $c_2$. For a given value of $c_1$ and $c_2$, we generate 50 scale-free networks using the preferential attachment algorithm and run the stochastic disease network game for $200$ steps for each network. Each run starts with all individuals infected. The grid color represents the average of infectivity level at time $t=200$ among runs. Increasing $c_1$ reduces the average number of infected in the population whether the disease is endemic or not. 
}
\label{fig:infectivity_level_behavior_beta}
\end{figure*}

See Figure \ref{fig:infectivity_level_behavior_beta} for the effect of parameters $c_1$, $c_2$, $\beta$ on infectivity level. 

\section{Epidemic threshold when $c_2 = 0$}\label{epidemic_threshold_without_empathy}

If $c_2 = 0$, an MMPE action profile at time $t$ can simply be written as follows,
\begin{align} \label{eq:MMPE_action_c2_0}
a_i^{*}(t) = \bbone\left(c_0 >  c_1 (1-s_i(t)) \sum_{j\in\ccalN_i} s_j(t) \right).
\end{align}
That is, the infected will socialize normally at all times and a susceptible agent will socialize if the number infected around is less than $\frac{c_0}{ c_1}$. Given the MMPE profile above, the infection probability of a susceptible individual $i$ is given by
\begin{align}
p^i_{01}(t) = P(s_i(t+1) = 1 | s_i(t) = 0)= 1 - \prod_{j\in\ccalN_i} (1-\beta a_i^{*}(t)) s_j(t).
\end{align}
We can bound the above conditional probability as follows by \eqref{eq:upper_bound_transition},
\begin{align} 
p^i_{01}(t) \leq  \beta a_i^{*}(t) \sum_{j\in\ccalN_i} s_j(t).
\end{align}

Next we obtain a $n$-state approximation of the Markov chain dynamics. The approximation follows the same steps in \cite{Paarporn_et_al_15}. Define the probability of infection of individual $i$ at time $t$ $p_i(t) = E[s_i(t)]$. Let $p(t)\in [0,1]^n$ be $n$ infection probability vector of the population.  By law of total probability, 
\begin{equation}
p_i(t+1) = p_i(t)(1-\delta) + (1 - p_i(t))p_{01}^i(t).
\end{equation}
Using the upper bound above for $p^i_{01}(t)$ and noting that $a_i^{*} = \bbone(c_0 > \beta c_1 \sum_{j\in\ccalN_i} s_j(t))$ from \eqref{eq:MMPE_action_c2_0} if $s_i(t) = 0$ , we obtain 
\begin{equation} \label{eq:sum_bound}
p_i(t+1) \leq p_i(t)(1-\delta) + (1 - p_i(t))\beta  \bbone\bigg(c_0 >  c_1\sum_{j\in\ccalN_i} s_j(t) \bigg)\sum_{j\in \ccalN_i} s_j(t).
\end{equation}
We now approximate the above bound by replacing $s_j(t)$ terms with $p_j(t)$ to get
\begin{equation} \label{eq:infection_probability_approximate}
p_i(t+1) \approx p_i(t)(1-\delta) + (1 - p_i(t))\beta \bbone\bigg(c_0 >  c_1\sum_{j\in\ccalN_i} p_j(t) \bigg) \sum_{j\in \ccalN_i} p_j(t).
\end{equation}

When we linearize the dynamics of $p(t)$  defined by \eqref{eq:infection_probability_approximate} around the fixed point origin $0_n$, i.e., the state of disease eradication $p_i(t) = 0$ for all $i$, we get the following,
\begin{equation} \label{eq:linear_disease_dynamics}
p(t+1) = ((1-\delta) I_n + \beta A ) p(t)
\end{equation}
where $I_n$ is the $n\times n$ identity matrix and $A \in [0,1]^{n\times n}$ is the adjacency matrix of the contact network, i.e., its $ij$th element $A_{ij} =1$ if $j \in \ccalN_i$, otherwise $A_{ij} = 0$. The approximate disease dynamics in \eqref{eq:linear_disease_dynamics} is linear. Furthermore, the origin $0_n$ is a globally stable fixed point if and only if $\lambda_{max} (  (1-\delta) I_n + \beta A) <1$ \cite{Mieghem_et_al_2009}. We can equivalently write this condition as $\frac{\beta}{\delta} \lambda_{max} (A) <1$. This result implies that the state of disease eradication is unstable if $\frac{\beta}{\delta} \lambda_{max} (A) >1$. Note that this threshold does not depend on the risk averseness constant $c_1$. This implies that risk aversion cannot by stop an outbreak when $c_2=0$, except in the extreme case of $c_0<c_1$.

\bibliographystyle{naturemag}
\bibliography{intro_bib}

\begin{thebibliography}{10}
\expandafter\ifx\csname url\endcsname\relax
  \def\url#1{\texttt{#1}}\fi
\expandafter\ifx\csname urlprefix\endcsname\relax\def\urlprefix{URL }\fi
\providecommand{\bibinfo}[2]{#2}
\providecommand{\eprint}[2][]{\url{#2}}

\bibitem{Pandey_et_al_2014}
\bibinfo{author}{Pandey, A.} \emph{et~al.}
\newblock \bibinfo{title}{Strategies for containing {Ebola} in {West}
  {Africa}}.
\newblock \emph{\bibinfo{journal}{Science}} \textbf{\bibinfo{volume}{346}},
  \bibinfo{pages}{991--995} (\bibinfo{year}{2014}).

\bibitem{Chowell_et_al_2003}
\bibinfo{author}{Chowell, G.}, \bibinfo{author}{Fenimore, P.~W.},
  \bibinfo{author}{Castillo-Garsow, M.~A.} \& \bibinfo{author}{Castillo-Chavez,
  C.}
\newblock \bibinfo{title}{{SARS} outbreaks in {Ontario}, {Hong} {Kong} and
  {Singapore}: the role of diagnosis and isolation as a control mechanism}.
\newblock \emph{\bibinfo{journal}{Journal of Theoretical Biology}}
  \textbf{\bibinfo{volume}{224}}, \bibinfo{pages}{1--8} (\bibinfo{year}{2003}).

\bibitem{Lau_et_al_2004}
\bibinfo{author}{Lau, J.~T.}, \bibinfo{author}{Tsui, H.}, \bibinfo{author}{Lau,
  M.} \& \bibinfo{author}{Yang, X.}
\newblock \bibinfo{title}{{SARS} transmission, risk factors, and prevention in
  {Hong} {Kong}}.
\newblock \emph{\bibinfo{journal}{Emerging Infectious Diseases}}
  \textbf{\bibinfo{volume}{10}}, \bibinfo{pages}{587--92}
  (\bibinfo{year}{2004}).

\bibitem{Pang_et_al_2003}
\bibinfo{author}{Pang, X.} \emph{et~al.}
\newblock \bibinfo{title}{Evaluation of control measures implemented in the
  severe acute respiratory syndrome outbreak in {Beijing}, 2003}.
\newblock \emph{\bibinfo{journal}{JAMA}} \textbf{\bibinfo{volume}{290}},
  \bibinfo{pages}{3215--3221} (\bibinfo{year}{2003}).

\bibitem{Hethcote_Yorke_1984}
\bibinfo{author}{Hethcote, H.~W.} \& \bibinfo{author}{Yorke, J.~A.}
\newblock \emph{\bibinfo{title}{Gonorrhea transmission dynamics and control}}
  (\bibinfo{publisher}{Springer Lecture Notes in Biomathematics},
  \bibinfo{year}{1984}).

\bibitem{Hyman_Li_1997}
\bibinfo{author}{Hyman, J.~M.} \& \bibinfo{author}{Li, J.}
\newblock \bibinfo{title}{Behavior changes in {SIS} {STD} models with selective
  mixing}.
\newblock \emph{\bibinfo{journal}{SIAM Journal on Applied Mathematics}}
  \textbf{\bibinfo{volume}{57}}, \bibinfo{pages}{1082--1094}
  (\bibinfo{year}{1997}).

\bibitem{Nelson_2004}
\bibinfo{author}{Nelson, R.~J.}
\newblock \bibinfo{title}{Seasonal immune function and sickness responses}.
\newblock \emph{\bibinfo{journal}{Trends in Immunology}}
  \textbf{\bibinfo{volume}{25}}, \bibinfo{pages}{187--192}
  (\bibinfo{year}{2004}).

\bibitem{Jones_Salathe_2009}
\bibinfo{author}{Jones, J.~H.} \& \bibinfo{author}{Salathe, M.}
\newblock \bibinfo{title}{Early assessment of anxiety and behavioral response
  to novel swine-origin influenza a ({H1N1})}.
\newblock \emph{\bibinfo{journal}{PLoS One}} \textbf{\bibinfo{volume}{4}},
  \bibinfo{pages}{e8032} (\bibinfo{year}{2009}).

\bibitem{Steelfisher_et_al_2010}
\bibinfo{author}{Steelfisher, G.~K.}, \bibinfo{author}{Blendon, R.~J.},
  \bibinfo{author}{Bekheit, M.~M.} \& \bibinfo{author}{Lubell, K.}
\newblock \bibinfo{title}{The public's response to the 2009 {H1N1} influenza
  pandemic}.
\newblock \emph{\bibinfo{journal}{New England Journal of Medicine}}
  \textbf{\bibinfo{volume}{362}}, \bibinfo{pages}{e65} (\bibinfo{year}{2010}).

\bibitem{Volz_Meyers_2007}
\bibinfo{author}{Volz, E.} \& \bibinfo{author}{Meyers, L.~A.}
\newblock \bibinfo{title}{Susceptible--infected--recovered epidemics in dynamic
  contact networks}.
\newblock \emph{\bibinfo{journal}{Proc. of the Royal Society of London B:
  Biological Sciences}} \textbf{\bibinfo{volume}{274}},
  \bibinfo{pages}{2925--2934} (\bibinfo{year}{2007}).

\bibitem{Meyers_et_al_2005}
\bibinfo{author}{Meyers, L.~A.}, \bibinfo{author}{Pourbohloul, B.},
  \bibinfo{author}{Newman, M. E.~J.}, \bibinfo{author}{Skowronski, D.~M.} \&
  \bibinfo{author}{Brunham, R.~C.}
\newblock \bibinfo{title}{Network theory and {SARS}: predicting outbreak
  diversity}.
\newblock \emph{\bibinfo{journal}{Journal of Theoretical Biology}}
  \textbf{\bibinfo{volume}{232}}, \bibinfo{pages}{71--81}
  (\bibinfo{year}{2005}).

\bibitem{Bansal_et_al_2007}
\bibinfo{author}{Bansal, S.}, \bibinfo{author}{Grenfell, B.~T.} \&
  \bibinfo{author}{Meyers, L.~A.}
\newblock \bibinfo{title}{When individual behaviour matters: homogeneous and
  network models in epidemiology}.
\newblock \emph{\bibinfo{journal}{Journal of the Royal Society Interface}}
  \textbf{\bibinfo{volume}{4}}, \bibinfo{pages}{879--891}
  (\bibinfo{year}{2007}).

\bibitem{Volz_Meyers_2009}
\bibinfo{author}{Volz, E.} \& \bibinfo{author}{Meyers, L.~A.}
\newblock \bibinfo{title}{Epidemic thresholds in dynamic contact networks}.
\newblock \emph{\bibinfo{journal}{Journal of the Royal Society Interface}}
  \textbf{\bibinfo{volume}{6}}, \bibinfo{pages}{233--241}
  (\bibinfo{year}{2009}).

\bibitem{Mieghem_et_al_2009}
\bibinfo{author}{Van~Mieghem, P.}, \bibinfo{author}{Omic, J.} \&
  \bibinfo{author}{Kooij, R.}
\newblock \bibinfo{title}{Virus spread in networks}.
\newblock \emph{\bibinfo{journal}{IEEE/ACM Transactions on Networking}}
  \textbf{\bibinfo{volume}{17}}, \bibinfo{pages}{1--14} (\bibinfo{year}{2009}).

\bibitem{Volz_et_al_2011}
\bibinfo{author}{Volz, E.~M.}, \bibinfo{author}{Miller, J.~C.},
  \bibinfo{author}{Galvani, A.} \& \bibinfo{author}{Meyers, L.~A.}
\newblock \bibinfo{title}{Effects of heterogeneous and clustered contact
  patterns on infectious disease dynamics}.
\newblock \emph{\bibinfo{journal}{PLoS Computational Biology}}
  \textbf{\bibinfo{volume}{7}}, \bibinfo{pages}{e1002042}
  (\bibinfo{year}{2011}).

\bibitem{Bauch_Galvani_2013}
\bibinfo{author}{Bauch, C.~T.} \& \bibinfo{author}{Galvani, A.~P.}
\newblock \bibinfo{title}{Epidemiology. {Social} factors in epidemiology.}
\newblock \emph{\bibinfo{journal}{Science (New York, NY)}}
  \textbf{\bibinfo{volume}{342}}, \bibinfo{pages}{47--49}
  (\bibinfo{year}{2013}).

\bibitem{Perra_et_al_2011}
\bibinfo{author}{Perra, N.}, \bibinfo{author}{Balcan, D.},
  \bibinfo{author}{Gon{\c{c}}alves, B.} \& \bibinfo{author}{Vespignani, A.}
\newblock \bibinfo{title}{Towards a characterization of behavior-disease
  models}.
\newblock \emph{\bibinfo{journal}{PloS One}} \textbf{\bibinfo{volume}{6}},
  \bibinfo{pages}{e23084} (\bibinfo{year}{2011}).

\bibitem{Mbah_et_al_2012}
\bibinfo{author}{Mbah, M. L.~N.} \emph{et~al.}
\newblock \bibinfo{title}{The impact of imitation on vaccination behavior in
  social contact networks}.
\newblock \emph{\bibinfo{journal}{PLoS Computational Biology}}
  \textbf{\bibinfo{volume}{8}}, \bibinfo{pages}{e1002469}
  (\bibinfo{year}{2012}).

\bibitem{Paarporn_et_al_15}
\bibinfo{author}{Paarporn, K.}, \bibinfo{author}{Eksin, C.},
  \bibinfo{author}{Weitz, J.~S.} \& \bibinfo{author}{Shamma, J.~S.}
\newblock \bibinfo{title}{Epidemic spread over networks with agent awareness
  and social distancing.}
\newblock In \emph{\bibinfo{booktitle}{{Proceedings of the 53rd Annual Allerton
  Conference on Communications, Control, and Computing (to appear)}}}
  (\bibinfo{address}{Allerton, Illinois, USA}, \bibinfo{year}{2015}).

\bibitem{Funk_et_al_2009}
\bibinfo{author}{Funk, S.}, \bibinfo{author}{Erez, G.},
  \bibinfo{author}{Watkins, C.} \& \bibinfo{author}{Jansen, V. A.~A.}
\newblock \bibinfo{title}{{The spread of awareness and its impact on epidemic
  outbreaks}}.
\newblock \emph{\bibinfo{journal}{Proceedings of The National Academy of
  Sciences USA}} \textbf{\bibinfo{volume}{106}}, \bibinfo{pages}{6872--6877}
  (\bibinfo{year}{2009}).

\bibitem{Bauch_et_al_03}
\bibinfo{author}{Bauch, C.~T.}, \bibinfo{author}{Galvani, A.~P.} \&
  \bibinfo{author}{Earn, D. J.~D.}
\newblock \bibinfo{title}{Group interest versus self-interest in smallpox
  vaccination policy}.
\newblock \emph{\bibinfo{journal}{Proc. of the National Academy of Sciences
  USA}} \textbf{\bibinfo{volume}{100}}, \bibinfo{pages}{10564--10567}
  (\bibinfo{year}{2003}).

\bibitem{Bauch_Earn_2004}
\bibinfo{author}{Bauch, C.~T.} \& \bibinfo{author}{Earn, D. J.~D.}
\newblock \bibinfo{title}{Vaccination and the theory of games}.
\newblock \emph{\bibinfo{journal}{Proc. of the National Academy of Sciences
  USA}} \textbf{\bibinfo{volume}{101}}, \bibinfo{pages}{13391--13394}
  (\bibinfo{year}{2004}).

\bibitem{Molina_Earn_2015}
\bibinfo{author}{Molina, C.} \& \bibinfo{author}{Earn, D. J.~D.}
\newblock \bibinfo{title}{Game theory of pre-emptive vaccination before
  bioterrorism or accidental release of smallpox}.
\newblock \emph{\bibinfo{journal}{Journal of The Royal Society Interface}}
  \textbf{\bibinfo{volume}{12}}, \bibinfo{pages}{20141387}
  (\bibinfo{year}{2015}).

\bibitem{Perisic_Bauch_2009}
\bibinfo{author}{Perisic, A.} \& \bibinfo{author}{Bauch, C.~T.}
\newblock \bibinfo{title}{Social contact networks and disease eradicability
  under voluntary vaccination}.
\newblock \emph{\bibinfo{journal}{PLoS Computational Biology}}
  \textbf{\bibinfo{volume}{5}}, \bibinfo{pages}{e1000280}
  (\bibinfo{year}{2009}).

\bibitem{Omic_et_al_2009}
\bibinfo{author}{Omic, J.}, \bibinfo{author}{Orda, A.} \&
  \bibinfo{author}{Van~Mieghem, P.}
\newblock \bibinfo{title}{Protecting against network infections: a game
  theoretic perspective}.
\newblock In \emph{\bibinfo{booktitle}{IEEE INFOCOM}},
  \bibinfo{pages}{1485--1493} (\bibinfo{year}{2009}).

\bibitem{Shim_et_al_2012}
\bibinfo{author}{Shim, E.}, \bibinfo{author}{Chapman, G.~B.},
  \bibinfo{author}{Townsend, J.~P.} \& \bibinfo{author}{Galvani, A.~P.}
\newblock \bibinfo{title}{The influence of altruism on influenza vaccination
  decisions}.
\newblock \emph{\bibinfo{journal}{Journal of The Royal Society Interface}}
  \bibinfo{pages}{rsif20120115} (\bibinfo{year}{2012}).

\bibitem{Enright_Kao_2015}
\bibinfo{author}{Enright, J.} \& \bibinfo{author}{Kao, R.~R.}
\newblock \bibinfo{title}{A few bad apples: A model of disease influenced agent
  behaviour in a heterogeneous contact environment}.
\newblock \emph{\bibinfo{journal}{PloS One}} \textbf{\bibinfo{volume}{10}},
  \bibinfo{pages}{e0118127} (\bibinfo{year}{2015}).

\bibitem{Zhang_et_al_13}
\bibinfo{author}{Zhang, H.-F.}, \bibinfo{author}{Yang, Z.},
  \bibinfo{author}{Wu, Z.-X.}, \bibinfo{author}{Wang, B.-H.} \&
  \bibinfo{author}{Zhou, T.}
\newblock \bibinfo{title}{Braess's paradox in epidemic game: better condition
  results in less payoff}.
\newblock \emph{\bibinfo{journal}{Scientific Reports}}
  \textbf{\bibinfo{volume}{3}} (\bibinfo{year}{2013}).

\bibitem{Cornforth_et_al_2011}
\bibinfo{author}{Cornforth, D.~M.} \emph{et~al.}
\newblock \bibinfo{title}{Erratic flu vaccination emerges from short-sighted
  behavior in contact networks}.
\newblock \emph{\bibinfo{journal}{PLoS Computational Biology}}
  \textbf{\bibinfo{volume}{7}}, \bibinfo{pages}{e1001062}
  (\bibinfo{year}{2011}).

\bibitem{Reluga_2010}
\bibinfo{author}{Reluga, T.~C.}
\newblock \bibinfo{title}{Game theory of social distancing in response to an
  epidemic}.
\newblock \emph{\bibinfo{journal}{PLoS Computational Biology}}
  \textbf{\bibinfo{volume}{6}}, \bibinfo{pages}{e1000793}
  (\bibinfo{year}{2010}).

\bibitem{Wang_et_al_2015}
\bibinfo{author}{Wang, Z.}, \bibinfo{author}{Andrews, M.~A.},
  \bibinfo{author}{Wu, Z.-X.}, \bibinfo{author}{Wang, L.} \&
  \bibinfo{author}{Bauch, C.~T.}
\newblock \bibinfo{title}{Coupled disease--behavior dynamics on complex
  networks: a review}.
\newblock \emph{\bibinfo{journal}{Physics of Life Reviews}}
  \textbf{\bibinfo{volume}{15}}, \bibinfo{pages}{1--29} (\bibinfo{year}{2015}).

\bibitem{Fudenberg_Tirole_1991}
\bibinfo{author}{Fudenberg, D.} \& \bibinfo{author}{Tirole, J.}
\newblock \emph{\bibinfo{title}{Game theory}} (\bibinfo{publisher}{The MIT
  Press, Cambridge, Massachusetts}, \bibinfo{year}{1998}),
  \bibinfo{edition}{6.} edn.

\bibitem{Mailath_Samuelson_2006}
\bibinfo{author}{Mailath, G.~J.} \& \bibinfo{author}{Samuelson, L.}
\newblock \emph{\bibinfo{title}{Repeated games and reputations}}
  (\bibinfo{publisher}{Oxford University Press, Oxford}, \bibinfo{year}{2006}),
  \bibinfo{edition}{2.} edn.

\bibitem{Dubey_1986}
\bibinfo{author}{Dubey, P.}
\newblock \bibinfo{title}{Inefficiency of {Nash} equilibria}.
\newblock \emph{\bibinfo{journal}{Mathematics of Operations Research}}
  \textbf{\bibinfo{volume}{11}}, \bibinfo{pages}{1--8} (\bibinfo{year}{1986}).

\bibitem{Keeling_Rohani_2011}
\bibinfo{author}{Keeling, M.} \& \bibinfo{author}{Rohani, P.}
\newblock \emph{\bibinfo{title}{Modeling infectious diseases in humans and
  animals}} (\bibinfo{publisher}{Princeton University Press},
  \bibinfo{address}{Princeton, NJ}, \bibinfo{year}{2011}).

\bibitem{Newman_2010}
\bibinfo{author}{Newman, M.}
\newblock \emph{\bibinfo{title}{Networks: an introduction}}
  (\bibinfo{publisher}{Oxford University Press, Inc.}, \bibinfo{address}{New
  York, NY, USA}, \bibinfo{year}{2010}).

\bibitem{Barabasi_et_al_1999}
\bibinfo{author}{Barab{\'a}si, A.-L.} \& \bibinfo{author}{Albert, R.}
\newblock \bibinfo{title}{Emergence of scaling in random networks}.
\newblock \emph{\bibinfo{journal}{Science}} \textbf{\bibinfo{volume}{286}},
  \bibinfo{pages}{509--512} (\bibinfo{year}{1999}).

\bibitem{Funk_et_al_2015}
\bibinfo{author}{Funk, S.} \emph{et~al.}
\newblock \bibinfo{title}{Nine challenges in incorporating the dynamics of
  behaviour in infectious diseases models}.
\newblock \emph{\bibinfo{journal}{Epidemics}} \textbf{\bibinfo{volume}{10}},
  \bibinfo{pages}{21--25} (\bibinfo{year}{2015}).

\end{thebibliography}

\end{document}